\documentclass[11pt, letterpaper]{article}

\usepackage{amsmath}
\usepackage{amsfonts}
\usepackage{amssymb}
\usepackage{graphicx}
\usepackage{simplewick}

\pdfoutput=1

\usepackage{color}

\usepackage[noadjust]{cite}

\usepackage{amsmath, amssymb, graphics, epsfig, graphicx}
\usepackage{epsf}
\usepackage{epstopdf}
\usepackage {amssymb}
\newcommand{\nc}{\newcommand}
\nc{\ba}{\begin{eqnarray}}
\nc{\ea}{\end{eqnarray}}
\newcommand\be{\begin{equation}}
\newcommand\ee{\end{equation}}

\newcommand{\calR}{{\cal{R}}}

\newcommand{\calP}{{\cal{P}}}

\newcommand{\bea}{\begin{eqnarray}}
\newcommand{\eea}{\end{eqnarray}}

\newcommand{\im}{{ \mathrm{Im} }}

\newcommand{\bfx}{{\bf{x}}}
\newcommand{\bfq}{{\bf{q}}}
\newcommand{\bfk}{{\bf{k}}}
\newcommand{\bfp}{{\bf{p}}}

\begin{document}

\vspace{5mm}
\vspace{0.5cm}
\begin{center}

\def\thefootnote{\fnsymbol{footnote}}

{ \bf \Large{ Two-loop corrections in  power spectrum in\\ models of inflation with PBHs formation
 } }
\\[1cm]

{ Hassan Firouzjahi$\footnote{firouz@ipm.ir}$}\\[0.5cm]

{\small \textit{ School of Astronomy, Institute for Research in Fundamental Sciences (IPM) \\ P.~O.~Box 19395-5746, Tehran, Iran }}\\

\end{center}

\vspace{.8cm}

\hrule \vspace{0.3cm}

\begin{abstract}
We calculate the two-loop corrections in primordial  power spectrum in  models of single field inflation incorporating an intermediate USR phase employed  for PBHs formation. 
Among  the total eleven one-particle irreducible Feynman diagrams, we calculate the  corrections from the ``double scoop" two-loop diagram involving two vertices of quartic Hamiltonians. We demonstrate that the fractional two-loop correction in power spectrum  scales like the square of the fractional one-loop correction. We confirm our previous findings that the loop corrections become arbitrarily large  in the setup where  the transition from the intermediate USR  to the final slow-roll phase is very sharp. This suggests that in order for the analysis to be under  perturbative control against loop corrections, one requires a mild transition with a long enough relaxation period towards the final attractor phase. 

\end{abstract}
\vspace{0.5cm} \hrule
\def\thefootnote{\arabic{footnote}}
\setcounter{footnote}{0}
\newpage
\section{Introduction}
\label{intro}

There have been intense debates in recent literature on the nature of loop corrections in  single field models of inflation involving an intermediate ultra slow-roll (USR) phase \cite{Kristiano:2022maq, Kristiano:2023scm, Riotto:2023hoz, Riotto:2023gpm, Choudhury:2023vuj,  Choudhury:2023jlt,  Choudhury:2023rks, Choudhury:2023hvf, 
Choudhury:2024one, Choudhury:2024aji, Firouzjahi:2023aum, Motohashi:2023syh, Firouzjahi:2023ahg, Tasinato:2023ukp, Franciolini:2023agm, Firouzjahi:2023btw, Maity:2023qzw, Cheng:2023ikq, Fumagalli:2023loc, Nassiri-Rad:2023asg, Meng:2022ixx, Cheng:2021lif, Fumagalli:2023hpa,  Tada:2023rgp,  Firouzjahi:2023bkt, Iacconi:2023slv, Davies:2023hhn, Iacconi:2023ggt, Kristiano:2024vst, Kristiano:2024ngc, Ballesteros:2024zdp, Kawaguchi:2024lsw, Braglia:2024zsl, Firouzjahi:2024psd, Caravano:2024moy, Caravano:2024tlp, Saburov:2024und},	for   earlier works concerning the  quantum loop effects  in models of   inflation see  \cite{ Seery:2007wf, Seery:2007we, Senatore:2009cf, Pimentel:2012tw, Inomata:2022yte}. 
These models have been employed to generate primordial black holes (PBHs) as candidates for the observed dark matter \cite{Ivanov:1994pa, Garcia-Bellido:2017mdw, Germani:2017bcs, Biagetti:2018pjj}, for a review on the mechanism of  generating PBHs from USR setup see \cite{Khlopov:2008qy, Ozsoy:2023ryl, Byrnes:2021jka, Escriva:2022duf, Pi:2024jwt}.  In the USR setup involving  a flat potential, the  curvature perturbation grows on superhorizon scales. The enhancement	in	power spectrum allows one	to use this	setup to generate 
PBHs on desired scales.  However, the rapid growth of the  power spectrum during the intermediate USR stage can be problematic. 
More specifically, Kristiano and Yokoyama argued in  \cite{Kristiano:2022maq} that the  one-loop corrections originated from the small scales USR modes  can affect the  CMB perturbations. Correspondingly, it was concluded originally in \cite{Kristiano:2022maq} that the analysis are not under perturbative control and the setup is not trusted for PBHs formation.  Following	\cite{Kristiano:2022maq}, the  one-loop corrections in curvature perturbation power spectrum   were studied with different (conflicting) conclusions. 
For example,  the results of \cite{Kristiano:2022maq} was  criticized by Riotto  \cite{ Riotto:2023gpm, Riotto:2023hoz} arguing  that  the one-loop corrections can be small if the transition to the final attractor phase is smooth enough. Similarly, in  \cite{Firouzjahi:2023ahg},  employing $\delta N$ formalism,  it was argued that the large loop effects in the models with mild transitions  are suppressed by the slow-roll parameters and the model is under  perturbative control  for generating PBHs. In addition, the loop corrections  were studied numerically in \cite{Davies:2023hhn} and also employing  the formalism of separate universe in  \cite{Iacconi:2023ggt}.

In order to estimate the total one-loop corrections in  power spectrum, we need to have  both the cubic and quartic  Hamiltonians. The cubic action was calculated originally by Maldacena \cite{Maldacena:2002vr} but there is no concrete result for the full quartic interaction including the USR phase, but for earlier studies on quartic action see \cite{Jarnhus:2007ia, Arroja:2008ga}.  In \cite{Firouzjahi:2023aum}, we have employed the formalism of effective field theory (EFT) of inflation which enabled us to calculate the cubic and the quartic Hamiltonians in reasonable ease in the decoupling limit.  
Furthermore, we were able to incorporate the effects of the sharpness of the transition from  the USR stage to the final slow-roll phase too. We have shown  in \cite{Firouzjahi:2023aum} that  the one-loop corrections can be dangerous in models with a sharp transition, supporting the conclusion of \cite{Kristiano:2022maq}. 

Based on the physical intuitions and the expectation  of the decoupling of scales, it looks  counterintuitive that the  small USR modes can affect the long CMB scales  in the first place \cite{ Riotto:2023gpm, Firouzjahi:2023bkt}.  Motivated by this question,   the conclusion of  \cite{Kristiano:2022maq}  was critically revisited in \cite{Fumagalli:2023hpa} and \cite{Tada:2023rgp} where it was claimed that the one-loop corrections are canceled.  Specifically, in \cite{Fumagalli:2023hpa} the contributions of the boundary terms, which were not taken into account in previous works, were highlighted.   On the other hand, it was argued in \cite{Tada:2023rgp}   that the loop  contributions  vanish after  the  UV limit of the momentum is considered via some $i \varepsilon$ prescription. However, in both \cite{Fumagalli:2023hpa, Tada:2023rgp}, like many other previous works, only the cubic interactions were considered. The conclusions of \cite{Fumagalli:2023hpa, Tada:2023rgp} were reviewed critically in \cite{Firouzjahi:2023bkt} highlighting the flaws in their arguments. More recently, there were new  
claims of loop cancellation in \cite{Inomata:2024lud, Kawaguchi:2024rsv, Fumagalli:2024jzz}, we would like to come back to the results of these claims elsewhere. 

As for the physical origins of the loop corrections, the non-linear coupling between the short and long modes induces a source term in the equation governing the  evolution of the long mode. At the same time,  the spectrum of 
the short modes are modulated by the long modes.  The modulation effects becomes large if the power spectrum of the short modes is  highly scale-dependent which is the case  in the USR model.  The combined effects of the  non-linear coupling between the short  and long modes and the modulation of the short modes by the long mode back-reacts on the long mode itself, causing the loop corrections as highlighted  in  \cite{Riotto:2023hoz, Franciolini:2023agm, Firouzjahi:2023bkt}. 

In the light of the above discussions, in this work we aim to calculate the  corrections in power spectrum at the two-loop order. As one	may	expect, the analysis at two-loop order are significantly more complicated than in one-loop case. First, we have more Feynman diagrams involving not only the cubic and quartic interactions, but also the quintic and sextic Hamiltonians. Secondly, most of these Feynman diagrams involve double or higher order nested in-in integrals which make the time integral very complicated. As we shall see, there are in total eleven one-particle irreducible Feynman diagrams at two-loop level. Out of these eleven diagrams, we consider the case of the ``double scoop" digram which involves a double nested time integral containing two quartic Hamiltonians. We believe that	the results from this case are illustrative enough which can shed light on the structure of the two-loop corrections.


\section{The Setup}
\label{setup}

In this section we briefly review our setup. This is the same setup as employed in \cite{Kristiano:2022maq}  for PBHs formation.  It  is a  single field inflation with three distinct phases, $SR \rightarrow USR \rightarrow SR$, where the first and and the third stages are assumed to be  in SR  phases while the 
USR phase is sandwiched in between. The observed  CMB perturbations leave the horizon during the first  SR phase  with the amplitude of power spectrum set by the COBE normalization.  However, the intermediate USR phase,  engineered to produce the PBHs at the desired mass scales, may start at about 30 e-folds after the  CMB scales have left the Hubble horizon and typically it  lasts for 2-3 e-folds. Finally,  the intermediate USR phase is followed by the second SR phase where it is assumed that 
the system reaches its attractor stage.

As usual, during the SR phases the curvature perturbation $\calR$ is frozen 
after the mode leaves the horizon.  However, during the USR phase it experiences an exponential growth \cite{Kinney:2005vj, Namjoo:2012aa, Martin:2012pe, Chen:2013aj, Morse:2018kda, Lin:2019fcz, Dimopoulos:2017ged}. The rapid growth of the modes which become superhorizon 
 during the intermediate USR phase is the key idea behind the enhancement in curvature perturbation power spectrum to generate PBHs on the corresponding scales. In addition, the superhorizon growth of curvature perturbation plays important roles in  violating  the Maldacena non-Gaussianity consistency condition in USR model \cite{Namjoo:2012aa, Martin:2012pe, Chen:2013aj, Chen:2013eea, Akhshik:2015nfa, Akhshik:2015rwa, Mooij:2015yka, Bravo:2017wyw, Finelli:2017fml, Passaglia:2018ixg, Pi:2022ysn, Ozsoy:2021pws,  Firouzjahi:2023xke, Namjoo:2023rhq, Namjoo:2024ufv, Cai:2018dkf}.

Starting with the FLRW metric,
\ba
ds^2 = -dt^2 + a(t)^2 d{\bf x}^2 \, ,
\ea
the equations governing the dynamics of  the background  inflaton field $\phi$ and the scale factor $a(t)$ during the USR stage are,
\ba
\ddot \phi(t) + 3 H \dot \phi(t)=0\, , \quad \quad 3 M_P^2 H^2 \simeq V_0, 
\ea
in which $M_P$ is the reduced Planck mass, $H$ is the Hubble expansion  rate and $V_0$ is the  constant potential during the USR stage.   

Since  $H$ is very nearly constant during the USR phase then  
$\dot \phi \propto a^{-3}$.   Correspondingly, the first slow-roll parameter 
$\epsilon \equiv -\frac{\dot H}{H^2} $ falls off like $a^{-6}$ while the second slow-roll parameter $\eta \equiv \frac{\dot \epsilon}{H \epsilon}$ is nearly constant, $\eta\simeq-6$. It is assumed that  the USR phase is extended in the interval $\tau_s < \tau <\tau_e$ so  $\epsilon$ at the time of end of USR is given by $\epsilon_e = \epsilon_i \big( \frac{\tau_e}{\tau_s} \big)^6 $ where $\epsilon_i$ is the value of $\epsilon$ at the start of USR phase. 
Here $\tau$ is the conformal time which is related to cosmic time as usual via $d\tau= dt/a(t)$ with the understanding that towards the end of inflation $\tau \rightarrow 0$. 
Alternatively, working with  the number of e-fold  $d N= H dt$ as the clock, the duration of the USR  is determined by $\Delta N \equiv N(\tau_e) - N(\tau_s)$ 
yielding to    $\epsilon_e = \epsilon_i e^{-6 \Delta N}  $. For PBHs formation we typically require $\Delta N$ to be around 2 to 3 e-folds.  

To simplify the analysis, we assume the transitions  $SR \rightarrow USR$ and $USR \rightarrow SR$ happen instantaneously, at  $\tau=\tau_s$ and $\tau=\tau_e$ respectively. However, it may take time to end up in its attractor phase during the final SR phase. This is determined by the sharpness (actually the relaxation) parameter $h$ initially defined in \cite{Cai:2018dkf} via 
\ba
\label{h-def}
h
=-6 \sqrt{\frac{\epsilon_V}{\epsilon_e}} \, ,
\ea
in which $\epsilon_V$ is the value of $\epsilon$ during the final slow-roll phase which is determined as usual by the first derivative of the potential. Note that by construction $h<0$. In our analysis, we assume a sharp enough transition so $|h| >1$. For mild transition, say $h$ at the order of slow-roll parameters, the mode function keeps evolving during the final phase and the analysis become complicated. However, for a sharp transition, the system reaches the attractor phase quickly and the errors in our analytical results are expected to be negligible. 

For a very sharp transition with $h \rightarrow -\infty$,  $\epsilon$ approaches rapidly  to a larger value such that towards the end of inflation,  
$\epsilon(\tau_0) \simeq \epsilon_V = \epsilon_e (\frac{h}{6})^2$. 
On the other hand, for an ``instant" sharp transition which was assumed  in \cite{Kristiano:2022maq, Kristiano:2023scm}, one has $h=-6$.  In this situation  $\epsilon$ in the third SR phase is equal to its value at the time of end of USR, i.e. $\epsilon_V =\epsilon_e$. 

The evolution of the slow-roll parameters after the USR phase are studied in 
 \cite{Cai:2018dkf}. In particular, $\epsilon$ is  smooth  across the transition point but  $\eta$ experiences a jump  at $\tau=\tau_e$. Prior to the transition and close to the end of USR, $\eta=-6$ while right after the transition we have  $\eta= -6-h$. Following 
\cite{Cai:2018dkf}, we can approximate $\eta $ near the transition point as follows, 
\ba
\eta = -6 - h \theta(\tau -\tau_e) \quad \quad  \tau_e^- < \tau < \tau_e^+ \, ,
\ea
yielding to    
\ba
\label{eta-jump}
\frac{d \eta}{d \tau} = - h \delta (\tau -\tau_e)  \, ,  \quad \quad  \tau_e^- < \tau < \tau_e^+ \, .
\ea 
As we shall see, the jump in $\eta$ highlighted by the Dirac delta function above, plays crucial role in the loop corrections.


After presenting our background, we briefly review the perturbations in this setup.  To perform the in-in analysis we need the mode function associated to the  comoving curvature perturbation  $\calR$ during the USR and afterwards. In the Fourier space,  the mode function is
\begin{equation}
\calR({\bf x}, t) = \int \frac{d^3 k}{(2\pi)^3} e^{i {\bf k}\cdot {\bf x}} \hat\calR_{\bf k}(t) \, ,
\end{equation}
where, as usual,  the operator $\hat\calR_{\bf k}(t)$ is expressed in terms of the creation and annihilation operators $a_{\bf k}$ and $a_{\bf k}^\dagger$
via $\hat\calR_{\bf k}(t)= \calR_k(t) a_{\bf k} + \calR^*_k(t) a_{-\bf k}^\dagger$. In this notation,  $\hat\calR_{\bf k}$ is a quantum operator and 
 $\calR_k$ is the mode function.  As usual, the creation and annihilation operators satisfy the standard commutation relations $[ a_{\bf k}, a^\dagger_{\bf k'} ] = ( 2 \pi)^3 \delta (  {\bf k} - {\bf k'}) $.

The quantum initial condition is fixed by the  Bunch-Davies vacuum with the mode function, 
 \begin{equation}
\calR^{(1)}_{k} =  \frac{H}{ M_P\sqrt{4 \epsilon_i k^3}}
( 1+ i k \tau) e^{- i k \tau} \, , \quad \quad (\tau < \tau_s) \, .
\end{equation}
During the intermediate USR phase, the mode function is parameterized via,
\begin{equation}
\calR^{(2)}_{k} =  \frac{H}{ M_P\sqrt{4 \epsilon_i k^3}}  \bigg( \frac{\tau_s}{\tau} \bigg)^3
\Big[ \alpha^{(2)}_k ( 1+ i k \tau) e^{- i k \tau}  + \beta^{(2)}_k ( 1- i k \tau) e^{ i k \tau}  \Big]  \, ,
\end{equation}
in which the coefficients $\alpha^{(2)}_k$ and $\beta^{(2)}_k$ are determined by imposing the continuity of the mode function and its time derivative at $\tau =\tau_s$, yielding \begin{equation}
\label{alpha-beta2}
\alpha^{(2)}_k = 1 + \frac{3 i }{ 2 k^3 \tau_s^3} ( 1 + k^2 \tau_s^2) \, , \quad \quad
\beta^{(2)}_k= -\frac{3i }{ 2 k^3 \tau_s^3 } {( 1+ i k \tau_s)^2} e^{- 2 i k \tau_s} \, .
\end{equation}
On the other hand, imposing the matching conditions at $\tau_e$, the outgoing mode function during the final SR phase  is given by \cite{Firouzjahi:2023aum},
\begin{equation}
\calR^{(3)}_{k} =  \frac{H}{ M_P\sqrt{4 \epsilon(\tau) k^3}}
\Big[ \alpha^{(3)}_k ( 1+ i k \tau) e^{- i k \tau}  + \beta^{(3)}_k ( 1- i k \tau) e^{ i k \tau}  \Big] \, ,
\end{equation}
in which $\alpha^{(3)}_k$ and $\beta^{(3)}_k$ are determined to be,
$$
\label{alpha-beta3}
\alpha^{(3)}_k = \frac{1}{8 k^6 \tau_s^3 \tau_e^3}  \Big[ 3h
 ( 1 -i k \tau_e)^2 (1+i k \tau_s)^2 e^{2i k (\tau_e- \tau_s)}
-i (2 k^3 \tau_s^3 + 3i k^2 \tau_s^2 + 3 i) (4 i k^3 \tau_e^3- h k^2 \tau_e^2 - h) \Big],
\nonumber
$$
and
$$
\beta^{(3)}_k=   \frac{-1}{8 k^6 \tau_s^3 \tau_e^3}  \Big[ 3 ( 1+ i k \tau_s)^2 ( h+ h k^2 \tau_e^2 + 4 i k^3 \tau_e^3 ) e^{-2 i k \tau_s} + i h ( 1+ i k \tau_e)^2  ( 3 i + 3 i k^2 \tau_s^2 + 2 k^3 \tau_s^3 ) e^{- 2 i k \tau_e}
 \Big]. \nonumber
$$

With the mode functions given above, we can calculate the two-loop corrections in curvature perturbations power spectrum. We choose the convention that  the momentum associated to the long CMB modes are denoted by  ${\bf p}_1$ and   ${\bf p}_2$ while  the momentum corresponding to  the short modes which run inside the loops are denoted by ${\bf q}$ and ${\bf k}$. There is the vast hierarchy $p_i \ll q, k$.  In our analysis, we are interested in  the loop corrections induced from the short modes which become superhorizon  during the USR phase. Therefore,  we cut the momentum loop integrals in the range $ q_s \leq q < q_e $ where $q_s = -\frac{1}{\tau_s}$ and $q_e= - \frac{1}{\tau_e}$ are the modes which leave the horizon at $\tau=\tau_s$ and $\tau=\tau_e$ respectively.  Furthermore, the duration of USR period  $\Delta N\equiv  N(\tau_e) - N(\tau_s)$ is gievn in terms of  $q_s$ and $q_e$ by
\begin{equation}
\label{delta-N-def}
e^{- \Delta N}=  \frac{\tau_e}{\tau_s} = \frac{q_s}{q_e}  \, .
\end{equation}
As mentioned previously, to generate PBHs  with the desired mass scales, we require  $\Delta N \sim 2-3$.

In order to simplify the analysis, we have assumed an  instant transition at $\tau=\tau_e$ to the final SR phase. However, the mode functions keep evolving for $\tau > \tau_e$  before it assumes its final attractor value. This process  is governed by the relaxation parameter $h$. For example, for an instant and sharp transition which was employed in \cite{Kristiano:2022maq, Kristiano:2023scm} with $h =-6$,  $\calR$ at the end of inflation  is smaller by a factor of $1/4$  compared to its value at $\tau=\tau_e$.  The reason is that the mode function is not frozen right after the transition and it keeps evolving until it reaches its attractor value. However, for a very sharp transition corresponding to $h \rightarrow -\infty$, the mode function is assumed to freeze immediately after $\tau_e$. This is the situation considered in \cite{Namjoo:2012aa, Chen:2013aj} which yields  to $f_{NL}=\frac{5}{2}$. But, as demonstrated  in  \cite{Cai:2018dkf}, for a mild transition with $|h| \ll 1$,  non-Gaussianity is mostly erased during the second SR phase. 
Motivated by these discussions,  we should distinguish between an  instant transition and a sharp transition. For example, the assumption of an instant transition can be relaxed and one may  consider the situation where the transition to take place within some time interval \cite{Franciolini:2023agm, Davies:2023hhn}. However, this will complicate our theoretical analysis and is  beyond the scope of this work.


\section{Two-loops Feynman Diagrams and  Interaction Hamiltonians}

In order to obtain the loop corrections in curvature perturbation power spectrum  $\calP_\calR$, we need the interaction Hamiltonians. Here we present the structure of two-loops Feynman diagrams and the subset of interaction Hamiltonians necessary for our two-loop calculations.

To understand the structure of Feynman diagrams associated at two-loop level,  consider a general $L$-loop, one-particle irreducible Feynman diagram associated to the following scalar type potential
\ba
V= \sum_n g_n \phi^{n}  \, \quad \quad (n>2)\, ,
\ea
in which  $g_n$ is the coupling (vertex) and $n$ is the order of interaction. For example, for the cubic and  quartic interactions we have  $n=3$ and $n=4$ respectively.  Suppose we have a Feynman diagram with $L$ loops, 
$P$ internal propagators, $V_n$ vertices associated to each power of interaction $n$ and N external lines. For example, in our case of interest, $L=2$ (two-loops) and $N=2$ (two external lines for power spectrum).
Then using the following topological conditions \cite{Weinberg:1995mt},
\ba
L= P- \sum_n V_n + 1
\ea
and
\ba
N+ 2 P= \sum_n n V_n \, ,
\ea
we obtain the following relation between $L, N$ and $V_n$,
\ba
\label{relation1}
2 L= (2- N) + \sum_n (n-2) V_n \, .
\ea
For the loop corrections in power spectrum with $N=2$, this further simplifies to
\ba
\label{relation2}
2 L=  \sum_n (n-2) V_n  \,  \quad \quad (N=2) \, .
\ea
In particular, for one-loop corrections, the above condition allows only two 
Feynman diagrams, a single quartic vertex and a diagram with two cubic vertices as studied in details in \cite{Firouzjahi:2023aum}.

Now in our current case of interest with two-loop corrections ($L=2$), Eq. (\ref{relation2}) yields the following constraint,
\ba
4= V_3 + 2 V_4 + 3 V_5 + 4 V_6 \, .
\ea 
Based one the allowed  integer solutions of the above equation, one obtains the allowed Feynman diagrams. Here are all possible allowed solutions,
\ba
\label{sol-cons}
&(1)&: V_6= V_5=V_3=0,  ~ V_4=2\, ,  \nonumber\\
&(2)&: V_6= V_5= V_4=0, ~ V_3=4 ,  \nonumber\\
&(3)&: V_6= V_5=0, ~ V_4=1, V_3=2 ,  \nonumber\\
&(4)&: V_6=V_4=0,  ~  V_3=V_5=1 \, ,  \nonumber\\
&(5)&: V_3= V_4= V_5=0, ~ V_6=1 \,  .
\ea 

\begin{figure}[t]
\vspace{-.5 cm}
	\centering
	\includegraphics[ width=0.88\linewidth]{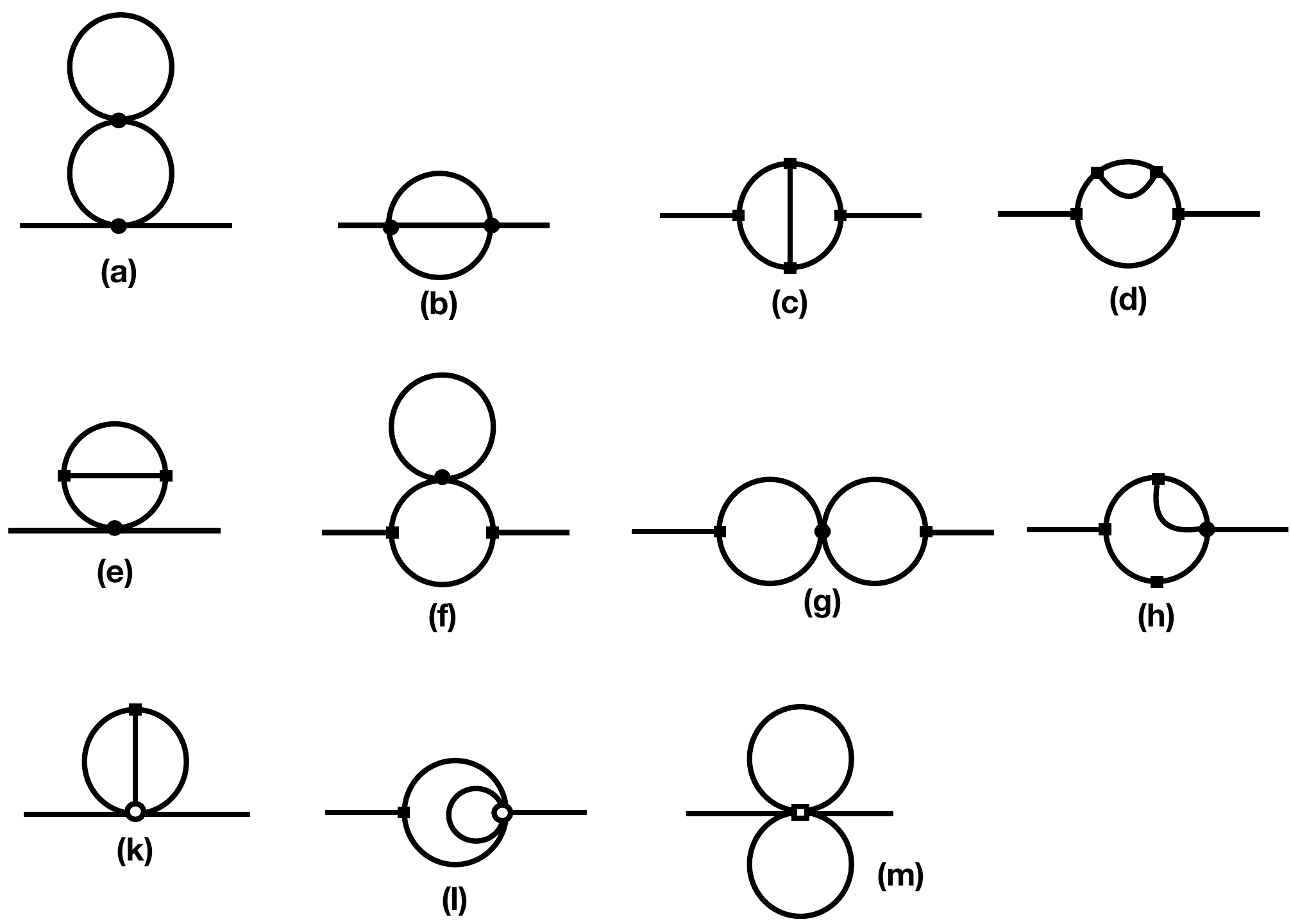}
	\vspace{.5 cm}
	\caption{ The one-particle irreducible Feynman diagrams for the two-loop corrections constructed from the solutions of Eq. (\ref{sol-cons}).  The diagrams $\bf{(a)}$ and $\bf{(b)}$ belong to category (1) in Eq. (\ref{sol-cons}), diagrams $\bf{(c)}$ and $\bf{(d)}$ to category (2), diagrams $\bf{(e), (f), (g)}$ and $\bf{(h)}$ to category (3), diagrams $\bf{(k)}$ and $\bf{(l)}$ belong to categories (4) while diagram 
	$\bf{(m)}$ belongs to category (5). }
\label{Feynman-fig}
\end{figure}

The above  solutions yield to 11 distinct two-loop diagrams as plotted in 
Fig. \ref{Feynman-fig}.  The Feynman diagrams in the first category involves two vertices of quartic Hamiltonian ${\bf H}_{4}$ (diagrams $\bf{(a)}$ and $\bf{(b)}$ in Fig. \ref{Feynman-fig}), while  the diagrams in the second category involve four vertices of the cubic Hamiltonian ${\bf H}_{3}$ (diagrams $\bf{(c)}$ and $\bf{(d)}$). The diagrams in the third category contain three vertices,  one from ${\bf H}_{4}$  and two from ${\bf H}_{3}$ (diagrams $\bf{(e)}$, $\bf{(f)}, \bf{(g)}$ and $\bf{\bf{(h)}}$). On the other hand, the diagram in the fourth category involves two vertices, one from  ${\bf H}_{3}$ and one from the quintic interaction ${\bf H}_{\bf 5}$ (diagrams $\bf{(k)}$ and $\bf{(l)}$).  Finally, the diagram in the fifth category involves a single vertex from the sextic Hamiltonian ${\bf H}_{\bf 6}$, (diagram $\bf{(m)}$).  
From the above discussions we see that we need ${\bf H}_{3}, {\bf H}_{4}, {\bf H}_{5}$ and ${\bf H}_{6}$ to calculate the full two-loop corrections in $\calP_\calR$. 

The cubic action for the curvature perturbations $\calR$ and the 
corresponding cubic Hamiltonian was calculated in details by Maldacena \cite{Maldacena:2002vr}. However, calculating the quartic action  
and the corresponding quartic Hamiltonian in this method is a very difficult task.  Fortunately, the formalism of EFT of inflation \cite{Cheung:2007st, Cheung:2007sv} provides a very useful alternative in which the interaction Hamiltonians can be calculated with reasonable ease. In particular, in the decoupling limit where the gravitational backreactions can be neglected, the EFT formalism was employed to calculate the cubic and quartic Hamiltonians in \cite{Firouzjahi:2023aum} (see also \cite{Akhshik:2015nfa} for the first work on this direction, calculating the cubic Hamiltonian). While the cubic and the quartic Hamiltonians were constructed in \cite{Firouzjahi:2023aum}, one still needs to calculate ${\bf H}_{5}$ and ${\bf H}_{6}$ to perform the full two-loop corrections. In principle, it is possible to calculate ${\bf H}_{5}$ and ${\bf H}_{6}$  using the EFT approach, but it turns out that there are new technical complications  which require careful considerations \cite{Nikbakht}.

The analysis of full two-loop corrections associated to the above 11 diagrams are a demanding task. As a first step forward, we calculate the two-loop corrections from the ``double scoop"  diagram $\bf{(a)}$ which is somewhat easier to handle technically. 
This is because this diagram involves two vertices so one deals with  double nested in-in integrals (this is also true for diagram $\bf{(b)}$). However, diagrams  $\bf{(c), (d), (e), (f), (g)}$ and $\bf{(h)}$ contain nested integrals with three-fold or four-fold time integrals involving ${\bf H}_{3}$ or ${\bf H}_{4}$
which are far more complicated than diagram $\bf{(a)}$.   As we shall see, the analysis even for the simple-looking diagram $\bf{(a)}$ is non-trivial. Having said this, physically one expects that the result obtained from this single diagram shed lights on the structure of two-loop corrections which should not be vary different than the remaining diagrams.

Here we briefly review the results of \cite{Firouzjahi:2023aum} which are required to calculate the quartic Hamiltonian to calculate the loop corrections from diagram $\bf{(a)}$.  We refer the reader to \cite{Firouzjahi:2023aum} for further details. 

The second order action employed to quantize the free theory is, 
\begin{equation}
\label{S2}
S_2= M_P^2 \int d\tau d^3 x\,    a^2 \epsilon H^2 \big(  \pi'^2 - (\partial_i \pi)^2 \big) \,,
\end{equation}
in which the prime represents the derivative with respect to $\tau$. Here $\pi(x^\mu)$ is the Goldstone boson associated to time diffeomorphism breaking which is related to curvature perturbations 
$\calR$ \cite{Cheung:2007st, Cheung:2007sv}.   

The cubic action is given by, 
\begin{equation}
\label{action3}
S_{\pi^3} =  M_P^2 H^3 \int d\tau d^3 x\,  \eta \epsilon  a^2\,
\Big[  \pi \pi'^2  - \pi (\partial \pi)^2 \Big]  \, ,
\end{equation}
and the corresponding  cubic interaction Hamiltonian is given by \cite{Akhshik:2015nfa},  
\begin{eqnarray}
\label{H3}
{\bf H}_3 =  
- M_P^2 H^3 \eta \epsilon a^2\, \int d^3 x\,   \Big[  \pi \pi'^2  +\frac{1}{2} \pi^2 \partial^2 \pi \Big] \, .
\end{eqnarray}

On the other hand,  the quartic action is obtained to be,  
\begin{equation}
\label{action4}
S_{\pi^4} =  \frac{M_P^2}{2} \int d\tau d^3 x\,   \epsilon a H^3
\big( \eta^2  a H + \eta' \big)\,    \Big[ \pi^2 \pi'^2 - \pi^2 (\partial \pi)^2 \Big] \, .
\end{equation}
In particular, note that there is  the  term $\eta' $ which induces the delta contribution  $\delta (\tau-\tau_e)$ in the interaction Hamiltonian when $\eta$ undergoes a jump at $\tau=\tau_e$ as given by Eq.~(\ref{eta-jump}).

As discussed in  \cite{Firouzjahi:2023aum}, in calculating the quartic Hamiltonian, care must be taken as the time derivative interaction $\pi' \pi^2$ in ${\bf H}_3$ induces an additional contribution in  the quartic Hamiltonian~\cite{Chen:2006dfn, Chen:2009bc}. As a result, one can not simply conclude that ${\bf H}_4 = -{\bf L}_4$.  More specifically, the quartic Hamiltonian receives additional  contribution  $+M_P^2 H^4  \eta^2 \epsilon a^2\,    \pi^2 \pi'^2$ from the cubic action. Combining all contributions, the total quartic Hamiltonian is given by \cite{Firouzjahi:2023aum}
\begin{equation}
\label{H4}
{\bf H}_4 =  \frac{M_P^2}{2} \epsilon a H^3  \int d^3 x \Big[
\big(  \eta^2  a H - \eta'   \big)  \pi^2 \pi'^2
+ \big(  \eta^2  a H + \eta'  \big)  \pi^2 (\partial \pi)^2
\Big] \, .
\end{equation}
The interaction Hamiltonians (\ref{H3}) and (\ref{H4}) have been used in 
\cite{Firouzjahi:2023aum} to calculate the one-loop corrections to power spectrum. We comment that in obtaining the above Hamiltonian, we have ignored total time derivatives in the form $\frac{d}{dt}\big(f(t) \pi^4\big)$ where 
$f(t)$ is a function of the background quantity. However, as shown in \cite{Braglia:2024zsl}, these boundary terms are harmless as they do not involve $\dot \pi$ and their contributions can be absorbed via a canonical transformation in phase space. 

The quantity of interest is the curvature perturbation $\calR$ while the above interaction Hamiltonians are given in terms of the Goldstone field $\pi$. The relation between $\calR$ and $\pi$ is non-linear.  For example,  to cubic order in $\pi$, they are related to each other via ~\cite{Jarnhus:2007ia, Arroja:2008ga}, 
 \begin{eqnarray}
 \label{pi-R1}
 \calR &=& - H \pi + \big( H \pi \dot \pi + \frac{\dot H}{2} \pi^2 \big)
 + \big( -H \pi \dot \pi^2 -\frac{H}{2} \ddot \pi \pi^2 - \dot H \dot \pi \pi^2 -\frac{\ddot H}{6} \pi^3 \big) \, , \nonumber\\
 &=& - H \pi + \frac{1}{2} \frac{d}{dt} \big(H \pi^2 \big) -  \frac{1}{6} \frac{d^2}{dt^2} \big(H \pi^3 \big) \, .
 \end{eqnarray}

However, we calculate the loop corrections in power spectrum at the time of 
end of inflation $\tau=\tau_0 \rightarrow 0$ when it is assumed that 
the system is in the slow-roll phase and the long mode perturbations are frozen with  $\dot \pi =\ddot \pi=0$. Fortunately, one can neglect the non-linear corrections in $\calR$ in Eq.~(\ref{pi-R1}) in this limit and  simply consider the  linear relation between them, 
 \begin{equation}
 \label{pi-R}
 \calR = - H \pi \, ,\quad  \quad \quad (\tau \rightarrow \tau_0).
 \end{equation}
Since the relation between   $\calR$ and $\pi$ is linear at $\tau_0$, we can simply write  $\langle\calR(\tau_0) \calR(\tau_0)\rangle= H^2 \langle\pi(\tau_0) \pi(\tau_0)\rangle$.  Consequently, one can
use $\pi$ and $\calR$ interchangeably in the following in-in analysis. More specifically, we will use the free mode function of $\calR$ in the interaction picture in place of the $\pi$ perturbations in the following in-in integrals.

\section{Loop Corrections in Power Spectrum }

Employing the perturbative in-in formalism \cite{Weinberg:2005vy}, 
the expectation value of the quantum operator $\hat {O}[\tau_0]$ measured at the time of end of inflation $\tau_0$ is, 
 \begin{equation}
 \label{Dyson}
 \langle \hat O(\tau_0) \rangle = \Big \langle \Big[ \bar {\mathrm{T}} \exp \Big( i \int_{-\infty}^{\tau_0} d \tau' H_{\mathrm{in}} (\tau') \Big) \Big] \,  \hat O(\tau_0)  \, \Big[ \mathrm{T} \exp \Big( -i \int_{-\infty}^{\tau_0} d \tau' H_{\mathrm{in}} (\tau') \Big) \Big]
 \Big \rangle \, .
 \end{equation}
 Here, as usual, $\mathrm{T}$ and $\bar {\mathrm{T}}$ represent the time ordering and anti-time ordering respectively and $H_{\mathrm{in}}(t)$ is the interaction Hamiltonian. For our case  of interest here $ \hat O(\tau_0)= \calR_{\bfp_1}(\tau_0) \calR_{\bfp_2}(\tau_0)$, while for the Feynman diagram $\bf{(a)}$ which we consider, we only need  the quartic interactions so  $H_{\mathrm{in}} =  {\bf H}_4$.

In order to calculate the two-loop corrections, one requires to expand the in-in formula Eq. (\ref{Dyson}) to second orders $H_{\mathrm{in}}$. For this purpose, it is more convenient to use the Weinberg commutator method  associated to  Eq. (\ref{Dyson}). To second order in $H_{\mathrm{in}} =  {\bf H}_4$, we obtain \cite{Weinberg:2005vy},
\ba
\label{Weinberg}
 \langle \hat O (\tau_0) \rangle &=& i^2 \int_{-\infty}^{\tau_0} d\tau_2 \int_{-\infty}^{\tau_2} d \tau_1 \Big \langle \Big[ H_{\mathrm{in}}(\tau_1) , \big[ H_{\mathrm{in}}(\tau_2), \hat O (\tau_0) \big] \,  \Big] \Big \rangle \, \\
 \label{Weinberg2}
 &=& 2\int_{-\infty}^{\tau_0} d\tau_2 \int_{-\infty}^{\tau_2} d \tau_1
 \mathrm{Re} \Big[ \big \langle {\bf H_4} (\tau_1)  \hat O (\tau_0)  {\bf H_4} (\tau_2) \big \rangle 
 - \big \langle {\bf H_4} (\tau_1) {\bf H_4} (\tau_2)  \hat O (\tau_0)   \big \rangle 
 \Big] \, ,
\ea
with $\hat O (\tau_0) \equiv \calR_{\bfp_1}(\tau_0) \calR_{\bfp_2}(\tau_0)$. 

Depending on the contractions of external leg operators $\hat O (\tau_0) $,
there are two distinct Feynman diagrams as shown by diagram $\bf{(a)}$ and $\bf{(b)}$ in Fig.  \ref{Feynman-fig}.  The digram $\bf{(a)}$  corresponds to the situation in which $\hat O (\tau_0) $ contracts only with
${\bf H_4} (\tau_2)$, with no contractions to ${\bf H_4} (\tau_1)$. 
On the other hand, the diagram $\bf{(b)}$ corresponds to the case where 
$\hat O (\tau_0) $ contracts jointly  with both ${\bf H_4} (\tau_2)$ and 
${\bf H_4} (\tau_1)$. AS mentioned before, as a first try for the two-loop corrections, in this work, we only consider the  ``double scoop" diagram $\bf{(a)}$. 

It turns out to be very convenient to decompose the expectation values in terms of sub-component Wick contractions. As an  example, consider 
$\big \langle {\bf H_4} (\tau_1)  \hat O (\tau_0)  {\bf H_4} (\tau_2) \big \rangle$ in the second line in Eq. (\ref{Weinberg2}). It involves three forms of  contractions: 
$\contraction[1ex]{}{bc}{eeem}
({\bf H_4}(\tau_1){\bf H_4}(\tau_2)$,  $\contraction[1ex]{}{bc}{eem}
(\hat O (\tau_0) {\bf H_4}(\tau_2)$ 
and  $\contraction[1ex]{}{bc}{eeem}
({\bf H_4}(\tau_1){\bf H_4}(\tau_1)$. Let us define
\ba
\contraction[1ex]{}{bc}{eeem}
({\bf H_4}(\tau_1){\bf H_4}(\tau_2) \equiv  h(\tau_1, \tau_2)  \, , \qquad 
\contraction[1ex]{}{bc}{eem}
(\hat O (\tau_0)  { \bf H_4}(\tau_2) \equiv  g(\tau_2)   \, , \qquad 
\contraction[1ex]{}{bc}{eeem}
({\bf H_4}(\tau_1) {\bf H_4}(\tau_1) \equiv  c(\tau_1)   \, .
\ea
With these definitions, considering both terms in the second line of Eq. (\ref{Weinberg2}), we obtain, 
\ba
\label{B-eq}
\langle \hat O (\tau_0) \rangle   = -4   \int_{-\infty}^{\tau_0} d\tau_2 \int_{-\infty}^{\tau_2} d \tau_1 
 \mathrm{Im} [ g(\tau_2)] \,  \mathrm{Im}\big[  c(\tau_1)  h(\tau_1 , \tau_2)
 \big]  \, .
\ea
Our job is now to calculate the functions $ g(\tau_2),  c(\tau_1)$ and  
$ h(\tau_1 , \tau_2)$ for various possible contractions.  

To perform the in-in integrals, we write ${\bf H}_4$ in Eq. (\ref{H4}) 
as follows (neglecting the non-linear relations between $\pi$ and $\calR$ as discussed before), 
\ba
\label{H4-AB}
{\bf H_4} =  A_4 (\tau) \int d^3 \bfx \, \calR^2 \calR'^2 + 
B_4(\tau)  \int d^3 \bfx \, \calR^2  (\partial  \calR)^2 \, ,
\ea
with
\ba
A_4(\tau)    \equiv  \frac{1}{2} M_P^2 \eta^2 \epsilon a^2 \Big(  1- \frac{h}{\eta^2} \delta (\tau -\tau_e)  \tau_e \Big) \, , \quad \quad 
B_4(\tau)    \equiv  \frac{1}{2} M_P^2 \eta^2 \epsilon a^2 \Big(  1+ \frac{h}{\eta^2} \delta (\tau -\tau_e)  \tau_e \Big) \, .
\ea
In particular, note  the term $\delta (\tau -\tau_e) $  appearing above, 
which is originated  from the term $\eta'$ in 
${\bf H_4}$, see   Eq. (\ref{eta-jump}). 

Depending on which term  in ${\bf H_4}$ is contracted with 
each other and with $\hat O (\tau_0)$, we will have four different contributions  as follows:
\ba
\langle \hat O (\tau_0) \rangle 
= {\langle \hat O \rangle }_{A_4 A_4}+ {\langle \hat O \rangle}_{A_4 B_4}+ {\langle \hat O \rangle}_{B_4 A_4}  + {\langle \hat O \rangle}_{B_4 B_4} \, ,
\ea
For example, for $ {\langle \hat O \rangle }_{A_4 A_4}$ we have

\ba
\label{B-AA}
{\langle \hat O \rangle }_{A_4 A_4}= \int_{-\infty}^{\tau_0} d \tau_2 \int_{-\infty}^{\tau_2} d \tau_1 A_4(\tau_1) A_4(\tau_2)  \int d^3 \bfx \int d^3  {\bf y}
\Big \langle  \Big[ \calR^2 \calR'^2 (\bfx, \tau_1) ,  \big[ \hat O (\tau_0)  ,  \calR^2 \calR'^2  ({\bf{y}}, \tau_2) \big] \Big]
 \Big \rangle \, .
\ea
Going to the Fourier space, this is cast into 

\ba
\label{B-AA}
{\langle \hat O \rangle }_{A_4 A_4}  =  \int_{-\infty}^{\tau_0} d \tau_2 \int_{-\infty}^{\tau_2} d \tau_1  A_4(\tau_1) A_4(\tau_2) 
\Big[ \prod_i^4 \int \frac{d^3 \bfq_i}{ (2 \pi)^3} (2 \pi)^3 \delta^3( \sum_i \bfq_i)\Big] 
\Big[\prod_j^4 \int \frac{d^3 \bfk_i}{(2 \pi)^3} (2 \pi)^3 \delta^3( \sum_i \bfk_i)
\Big]  \hspace{-1cm} \nonumber\\
 \times  \bigg \langle   \bigg[   \big( \hat\calR_{\bfq_1}   \hat\calR_{\bfq_2} \hat\calR'_{\bfq_3} \hat\calR'_{\bfq_4} \big) (\tau_1), \Big[
 \big( \hat\calR_{\bfp_1} \hat\calR_{\bfp_2}  \big) (\tau_0)  , \big( \hat\calR_{\bfk_1}  \hat\calR_{\bfk_2} \hat\calR'_{\bfk_3}   \hat\calR'_{\bfk_4} \big) (\tau_2)  \Big]  \bigg]  \bigg \rangle \, .
\ea
After performing the contractions and imposing the constraints from the delta functions, in the soft limit where $p\ll q, k$, we end up with the following form of two-loop integrals 

\ba
{\big\langle \calR_{\bfp_1} \calR_{\bfp_2}(\tau_0) \big\rangle }_{A_4 A_4}= (2 \pi)^3 \delta^3(\bfp_1 + \bfp_2)
\int_{-\infty}^{\tau_0} d\tau_2
 \int_{-\infty}^{\tau_2} d\tau_1 A_4(\tau_1) A_4( \tau_2) 
 \int \frac{d^3 \bfq}{(2 \pi)^3} \int \frac{d^3 \bfk}{(2 \pi)^3} 
 F(\tau_1, \tau_2; k, q), \nonumber
\ea
in which the function $F(\tau_1, \tau_2; k, q)$ is determined by 
different values of 
$ c(\tau_1),  g(\tau_2)$ and $ h(\tau_1, \tau_2)$. 

\begin{figure}[t]
\vspace{-3 cm}
	\centering
	\includegraphics[ width=0.48\linewidth]{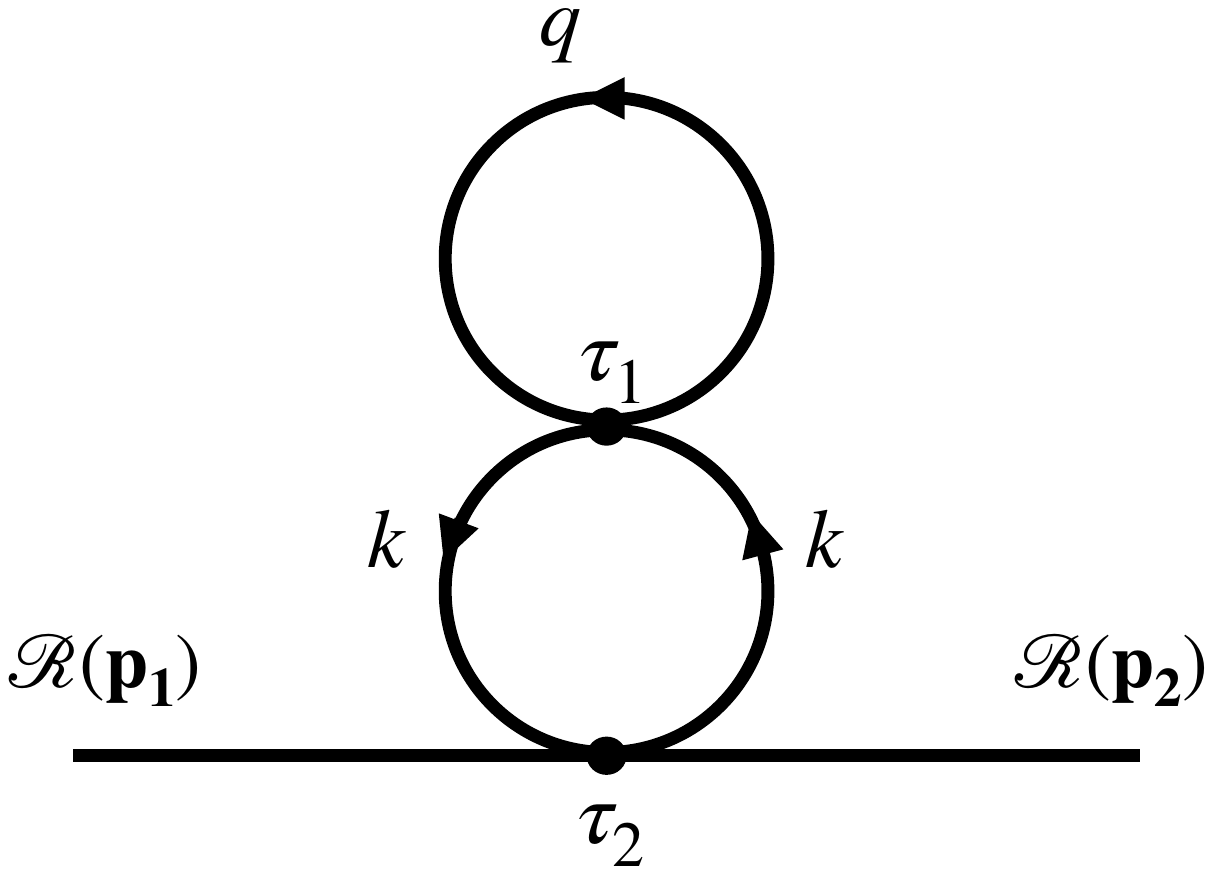}
	\vspace{-2 cm}
	\caption{ The arrangement of momenta inside each loop and the relative positions of $\tau_1$ and $\tau_2$.  }
\label{fig2}
\end{figure}


There are 9  different terms in $F(\tau_1, \tau_2; k, q)$  from different contractions. We list them as follows, 

\ba
\label{a0-term}
&&
(a_0):  (-4) (2)^2 \im \big[  \calR_p(\tau_0)^2  {\calR_p^{*}(\tau_2) }^2   \big]
\im\big[ {\calR_k'(\tau_1)}^2 {\calR_k^{'*}(\tau_2)}^2  \big] |\calR_q(\tau_1)|^2
\nonumber\\
&&
\label{b-term}
(b_0):  (-4) (2)^2 \im \big[  \calR_p(\tau_0)^2  {\calR_p^{*}(\tau_2) }^2   \big]
\im\big[ {\calR_k(\tau_1)}^2 {\calR_k^{'*}(\tau_2)}^2  \big] |\calR'_q(\tau_1)|^2
\nonumber\\
&&
\label{c0-term}
(c_0):  (-4) (2)^3 \im \big[  \calR_p(\tau_0)^2  {\calR_p^{*}(\tau_2) }^2   \big]
\im\big[ {\calR_k(\tau_1)} {\calR_k'(\tau_1)} {\calR_k^{'*}(\tau_2)}^2 \big]
\mathrm{Re} \big[\calR_q(\tau_1) \calR'_q(\tau_1)^*   \big],
\ea
and

\ba
\label{d0-term}
&&
(d_0):  (-4) (2)^4 \im \big[  \calR_p(\tau_0)^2  {\calR_p^{*}(\tau_2) } {\calR_p^{'*}(\tau_2) }   \big]
\im\big[ {\calR_k'(\tau_1)}^2 {\calR_k^{'*}(\tau_2)} {\calR_k^{*}(\tau_2)}  \big] |\calR_q(\tau_1)|^2
\nonumber\\
&&
\label{e-term}
(e_0):  (-4) (2)^4  \im \big[  \calR_p(\tau_0)^2  {\calR_p^{*}(\tau_2) } {\calR_p^{'*}(\tau_2) }   \big]
\im\big[ {\calR_k(\tau_1)}^2 {\calR_k^{'*}(\tau_2)} {\calR_k^{*}(\tau_2)}  \big] |\calR'_q(\tau_1)|^2
\\
&&
\label{f-term}
(f_0):  (-4) (2)^5  \im \big[  \calR_p(\tau_0)^2  {\calR_p^{*}(\tau_2) } {\calR_p^{'*}(\tau_2) }   \big]
\im\big[ {\calR_k(\tau_1)} {\calR_k'(\tau_1)} {\calR_k^{'*}(\tau_2)}  {\calR_k^{*}(\tau_2)} \big]
\mathrm{Re} \big[\calR_q(\tau_1) \calR'_q(\tau_1)^*   \big],  \nonumber
\ea
and

\ba
\label{m-term}
&&
(m):  (-4) (2)^2 \im \big[  \calR_p(\tau_0)^2  {\calR_p^{'*}(\tau_2) }^2   \big]
\im\big[ {\calR_k'(\tau_1)}^2 {\calR_k^{*}(\tau_2)}^2  \big] |\calR_q(\tau_1)|^2,
\nonumber\\
&&
\label{n-term}
(n):  (-4) (2)^2 \im \big[  \calR_p(\tau_0)^2  {\calR_p^{'*}(\tau_2) }^2   \big]
\im\big[ {\calR_k(\tau_1)}^2 {\calR_k^{*}(\tau_2)}^2  \big] |\calR'_q(\tau_1)|^2,
\nonumber\\
&&
\label{r0-term}
(r):  (-4) (2)^3 \im \big[  \calR_p(\tau_0)^2  {\calR_p^{'*}(\tau_2) }^2   \big]
\im\big[ {\calR_k(\tau_1)} {\calR_k'(\tau_1)} {\calR_k^{*}(\tau_2)}^2  \big]
\mathrm{Re} \big[\calR_q(\tau_1) \calR'_q(\tau_1)^*   \big] \, .
\ea
One subtle issue in the above expressions is the appearance of the term 
$\mathrm{Re} \big[\calR_q(\tau_1) \calR'_q(\tau_1)^*   \big]$ in terms $(c_0), (f_0)$ and $(r_0)$. This term originates from the contraction of $\hat\calR$ and $\hat\calR'$ in $\calR^2 \calR'^2$ in $A_4(\tau_1)$. As $\hat \calR$ and 
$\hat \calR'$ do not commute, we have symmetrized the ordering of 
$\hat \calR$ and  $\hat \calR'$ in $A_4(\tau_1)$ which yields to  $\mathrm{Re} \big[\calR_q(\tau_1) \calR'_q(\tau_1)^*   \big]$ in terms $(c_0), (f_0)$ and $(r_0)$. 

Looking at the above expressions we notice that the terms containing momentum $q$ are separated from the terms containing the momentum $k$.
This property is depicted in Fig. \ref{fig2} where the momentum $q$ runs in the top loop attached to time $\tau_1$ while the momentum $k$ runs in the lower loop attached to both $\tau_1$ and $\tau_2$. 
This is a simplifying feature of the diagram $\bf{(a)}$ in our Feynman diagrams. Because of this separation of the two momenta, it is technically easier to calculate the corrections from diagram $\bf{(a)}$ compared to diagram $\bf{(b)}$ in Fig.  \ref{Feynman-fig}.  While this is a simplification, but calculating the time integrals is still a non-trivial task. This is because we have two nested integrals over $\tau_1$ and $\tau_2$ involving 
10 factors of $\calR(\tau)$ and its derivatives.

The above three classes of contributions in Eqs. (\ref{c0-term}), (\ref{d0-term}) and (\ref{r0-term}) are grouped by their form of the function 
$ c(\tau_1)$ which depends only on the soft momentum $p$ but not on the loop momenta $q$ and $k$. Now, calculating these common factors in each class, we obtain
\ba
\label{first}
\im \big[  \calR_p(\tau_0)^2  {\calR_p^{*}(\tau_2) }^2   \big]= 
\frac{-H^4 \tau_s^6}{ 24 M_P^4 \epsilon_i^2 h \tau_e^3 \tau^3 p^3} 
\big(  h \tau_e^3 + (6-h) \tau^3\big)  \, ,
\ea
\ba
\label{second}
 \im \big[  \calR_p(\tau_0)^2  {\calR_p^{*}(\tau_2) } {\calR_p^{'*}(\tau_2) }   \big] 
 =\frac{H^4 \tau_s^6}{16 M_P^4 \epsilon_i^2 \tau^4 p^3} \, ,
\ea
and 
\ba
\label{third}
\im \big[  \calR_p(\tau_0)^2  {\calR_p^{'*}(\tau_2) }^2   \big]=
\frac{H^4 \tau_s^6 ( 6 \tau_s^5- \tau^5)}{40 M_P^4 \epsilon_i^2 \tau^8 p}\, .
\ea
Comparing the above three expressions, we conclude that the term in Eq. (\ref{third}) is much suppressed compared to terms Eq. (\ref{first}) and Eq. (\ref{second}) in the soft limit where $p \rightarrow 0$. Correspondingly, we can neglect the contributions of the terms $(m), (n)$ and $(r)$ in Eq. (\ref{r0-term}). 

We present the details of the analysis concerning the remaining three contributions ${\langle \hat O \rangle }_{A_4 B_4}, {\langle \hat O \rangle }_{B_4 A_4}$ and ${\langle \hat O \rangle }_{B_4 B_4}$ in the Appendix \ref{appendix}.  

Before presenting the final result, it is useful to have an estimate of the leading contributions. Let us look at the contributions of terms $(a_0)$ and $(b_0)$. The difference is that in $(a_0)$, $\calR'$ comes with $k$ inside the expression  $\im\big[ {\calR_k'(\tau_1)}^2 {\calR_k^{'*}(\tau_2)}^2  \big]$ while in $(b_0)$,  $\calR'$ involves the momentum $q$ simply as $|\calR'_q(\tau_1)|^2$.  On the other hand, for the superhorizon perturbations during the USR phase we have,
\ba
\label{R-prime}
\calR_q'(\tau) \simeq -\frac{3}{\tau} \calR_q(\tau) \propto \tau^{-4} \, .
\ea
Therefore, the leading effects in the in-in integrals on the superhorizon limit where $\tau_i \rightarrow 0$ are controlled by the contributions of  $\calR'$. 
Since the term $(b_0)$ involves  $|\calR'_q(\tau_1)|^2$, it scales like 
$\tau_1^{-8}$ while in $\im\big[ {\calR_k'(\tau_1)}^2 {\calR_k^{'*}(\tau_2)}^2  \big]$ the dependence can not be steeper than this. Indeed, calculating the leading terms on superhorizon limits of $\tau_1, \tau_2 \rightarrow 0$, one can show that the ratio $(a_0)/(b_0)$ scales like $(k \tau_1)^2$ which becomes smaller than unity on superhorizon scales. 
Therefore, it
is expected that the contribution of term $(b_0)$ to be more dominant than the term $(a_0)$. Indeed, calculating the in-in integrals for both terms $(a_0)$ and 
$(b_0)$, and neglecting the numerical prefactors and the common 
$(2 \pi)^3 \delta ^3(\bfp_1 + \bfp_2)$,  we obtain that they scale as follows:
\ba
(a_0): \large(\calP_{{\mathrm{CMB}}}\large)^3\,  e^{12 \Delta N} \Delta N \, , \quad \quad 
(b_0): \large(\calP_{{\mathrm{CMB}}}\large)^3\,  e^{12 \Delta N} \Delta N^2 \
\ea
in which $\calP_{{\mathrm{CMB}}}$ is the tree-level CMB scale power spectrum,
\ba
\label{P-CMB}
\calP_{{\mathrm{CMB}}}= \frac{H^2}{8 \pi^2 \epsilon_i M_P^2}\, .
\ea
For large enough $\Delta N$, the contributions from $(b_0)$ is typically larger than 
that of $(a_0)$ as expected.  

{As we demonstrated in  Appendix \ref{appendix}, in total there are 15 leading contributions (i.e. containing $p^{-3}$)  given by  $(a_0), (b_0), (c_0),$ $(d_0), (e_0), (f_0), (a_1), (b_1),$ $ (c_1),  (a_2), (b_2), (c_2), (d_2), (e_2), (f_2), (a_3)$ and $(b_3)$ presented in Appendix \ref{appendix}. 
 The contributions from these 15 terms 
are either like the contribution from the $(a_0)$ term scaling like $e^{12 \Delta N} \Delta N$ or like that of the $(b_0)$ term, scaling like $e^{12 \Delta N} \Delta N^2$. It turns out that only the contributions from terms $(b_0), (c_0), (d_0), (e_0)$ and $ (f_0)$ have the latter form.}

{In the limit of large enough $\Delta N$, and adding the contributions of terms $(b_0), (c_0), (d_0), (e_0)$ and $ (f_0)$ as the leading terms, we obtain the following fractional two-loop correction, }
\ba
\label{two-loop}
\frac{\Delta \calP^{(2-\mathrm{loop})}}{\calP_{{\mathrm{CMB}}}}\simeq  -\frac{27(23 h^2 + 132 h + 1152)}{8 h}\,  e^{12 \Delta N} \Delta N^2 \calP_{{\mathrm{CMB}}}^2 \, ,
\ea
where we have neglected the subleading terms involving $\Delta N e^{12 \Delta N}$. 
The fact hat the leading two-loop corrections scale like $e^{12 \Delta N} \Delta N^2 $ is both interesting and reassuring. In addition, the two-loop corrections scale linearly with $h$ for $h \rightarrow -\infty$. Note that our  result above is obtained assuming a  sharp enough transition with $|h| >1$ where  the effects of the relaxation during the final SR phase are neglected. 

It is instructive to compare the above two-loop corrections associated to diagram ({\bf{a}}) with the full one-loop correction obtained in \cite{Firouzjahi:2023aum},
\ba
\label{one-loop}
\frac{\Delta \calP^{(1-\mathrm{loop})}}{\calP_{{\mathrm{CMB}}}}
\simeq \frac{6(h^2 + 24 h + 180)}{ h} e^{6 \Delta N} \Delta N 
\calP_{{\mathrm{CMB}}}.
\ea
Comparing Eqs. (\ref{two-loop}) and (\ref{one-loop}), we obtain
\ba
\label{two-one}
\frac{\Delta \calP^{(2-\mathrm{loop})}}{\calP_{{\mathrm{CMB}}}}\sim
\Big( \frac{\Delta \calP^{(1-\mathrm{loop})}}{\calP_{{\mathrm{CMB}}}} \Big)^2 \, .
\ea
This is an interesting result, indicating that the fractional two-loop corrections 
is typically the square of the fractional one-loop correction.

For the loop effects to be under perturbative control, one requires that the successive loop corrections to be hierarchical, i.e. 
$\calP_{ {\mathrm{CMB}} } > \Delta \calP^{(1-\mathrm{loop})} > \Delta \calP^{(2-\mathrm{loop})}$. Neglecting the numerical prefactors,  from Eqs. (\ref{two-loop}) and (\ref{one-loop}) we obtain,
\ba
\frac{ \Delta \calP^{(2-\mathrm{loop})} }{ \Delta \calP^{(1-\mathrm{loop})}}
\sim  \frac{\Delta \calP^{(1-\mathrm{loop})}}{\calP_{{\mathrm{CMB}}}} \, \sim
e^{6 \Delta N} \Delta N  \calP_{{\mathrm{CMB}}} \, .
\ea
This shows that if the fractional one-loop correction is not small, then the two-loop corrections become significant as well so the perturbative treatment gets quickly out of control.
As elaborated in  \cite{Firouzjahi:2023aum}, the one-loop correction becomes significant for sharp transition when $|h| \gg1$ as can be seen in Eq. (\ref{one-loop}). Therefore, we conclude that the two-loop corrections become significant in model of sharp transition, which can be seen in Eq. (\ref{two-loop}) too. In order for the loop corrections to be under perturbative control, one has to either have a mild transition with $h$ at the order of  slow-roll parameters, or to take the duration of USR phase to be reasonably short, say $\Delta N \lesssim 1$. However, the latter arrangement may not be suitable for the PBHs formation as we need a long enough period of USR phase to enhance the power spectrum in the first place. For example, with $h=-6$, we need $\Delta N \lesssim 2.3$ to enhance $\calP_\calR$ by 7 orders of magnitudes compared to the CMB scale for the purpose of PBHs formation while still satisfying the perturbative 
bound from one-loop correction in Eq. (\ref{one-loop}). 
Therefore,  the only safe strategy for PBHs formation in this setup is to employ a mild transition with $|h| \ll 1$.

There are two important comments in order. The first comment is that in this work we studied the corrections from the diagram $\bf{(a)}$ 
in  Fig.  \ref{Feynman-fig}. Naturally, one may ask how the two-loop corrections from the remaining ten diagram can be compared to the current result given in Eq. (\ref{two-loop}). In a work in progress \cite{Nikbakht}, we are studying the correction from diagram ({\bf m}) in  Fig.  \ref{Feynman-fig} involving a single vertex of sextic Hamiltonian. We have confirmed that it scales like Eq. (\ref{two-loop}). On the physical ground, we expect the full two-loop corrections to have the same general form as in Eq. (\ref{two-loop}), i.e. scaling like $\Delta N^2 e^{12 \Delta N}$.  The second comment is the the issue of regularization and renormalization. In the current analysis, we have restricted the momentum in the range $q_s \leq q \leq k_e$. In principle one should integrate over the entire range $q_{IR} < q < q_{f}$ in which $q_{IR}$ is the lower IR limit while $q_{f}$ represents the mode which leaves the horizon 
at the time of end of inflation. Alternatively, one may simply set $q_{f} \rightarrow \infty$. While the IR contributions are under control, the UV contribution will diverge which has to be regularized and renormalized. Having said this, the peak of the power spectrum at the end of USR,  the dimensionless factor $e^{6 \Delta N} \calP_{{\mathrm{CMB}}}$, fixes the overall scale of the finite term after regularization. The fact that the two-loop correction in Eq. (\ref{two-loop}) is the square of the one-loop correction Eq. (\ref{one-loop}) qualitatively supports 
this expectation. However, it is an important open question to study the renormalization of the loop corrections in more detail. 

\section{Summary and Discussions } 
\label{Summary}

In this work, we have looked at the quantum corrections in primordial power spectrum at two-loop orders in models of single field inflation involving an intermediate  USR phase engineered for PBHs formation. 
This is the natural continuation of the previous works concerning the one-loop corrections. As the one-loop correction  in models with sharp transition to the final SR phase can be large \cite{Kristiano:2022maq, Firouzjahi:2023aum}, therefore it is necessary to examine the two-loop corrections for the severity of the loop contributions. As we have shown, there are 11 distinct one-particle irreducible Feynman diagrams at the two-loop orders. They require the cubic, quartic, quintic and sextic interaction Hamiltonians. 

As a first step forward, we have studied the corrections from the diagram {\bf (a)} in Fig. \ref{Feynman-fig}. This is because this diagram involves two vertices of quartic interaction while their momentum integrals over $q$ and $k$ are separable. This brings computational simplicities. Other diagrams in Fig. \ref{Feynman-fig} are more complicated, either having higher order nested integrals or 
the momentum integrals are not separable. On the other hand, the in-in analysis corresponding to diagram {\bf (m)} would be easier to handle as it involves a single sextic Hamiltonian vertex. However, one has to calculate the action to sixth order to construct ${\bf H_{6}}$ which involves additional technical complexities \cite{Nikbakht}. Our result shows that the two-loop corrections scale as the square of the one-loop corrections, i.e. like 
$\big(\Delta N e^{6 \Delta N}  \calP_{{\mathrm{CMB}}} \big)^2$. This is interesting and physically expected. 
This result confirms the previous results \cite{Kristiano:2022maq, Firouzjahi:2023aum}  that the loop corrections can quickly get out of control if the transition to the final attractor phase is very sharp and the duration of USR phase is long enough. In order for the loop corrections to be under perturbative control, and at the same time, to generate PBHs with the desired mass scales for dark matter purpose, it is necessary that the transition to the final attractor regime to be mild so the loop corrections are rendered harmless.
 
 The current work can be extended in various directions. One natural direction to proceed is to study the loop corrections from the remaining Feynman diagrams in Fig. \ref{Feynman-fig}. While this is an interesting and yet cumbersome  task, we believe that the total two-loop corrections would be similar to our current result, i.e. scaling like $\big(\Delta N e^{6 \Delta N} \calP_{{\mathrm{CMB}}}  \big)^2$. The other  direction to investigate is to study the general case of non-attractor inflation and the case of constant-roll inflation and see if the two-loop correction has a similar relation compared to the one-loop correction. Finally, an interesting direction to investigate is the question of regularization and renormalization  at both one-loop and two-loop orders. We would like to come back to this important question in future.


 \vspace{0.7cm}

 {\bf Acknowledgments:}  We thank Jacopo Fumagalli, 
 Haidar Sheikhahmadi, Amin Nassiri-Rad and Bahar Nikbakht  for useful discussions.  We thank ICCUB, University of Barcelona, for the kind hospitality during the workshop ``BBH initiative: I. Primordial Black Holes" where this work was in progress.  
This work is supported partially by INSF  of Iran under the grant  number 
4038049.

\appendix
\section{In-In Integrals }  
 \label{appendix}
In this Appendix we present the details of in-in analysis. 

As explained in the main text, the two-loop corrections have the following contributions,  \ba
\langle \hat O (\tau_0) \rangle 
= {\langle \hat O \rangle }_{A_4 A_4}+ {\langle \hat O \rangle}_{A_4 B_4}+ {\langle \hat O \rangle}_{B_4 A_4}  + {\langle \hat O \rangle}_{B_4 B_4} \, .
\ea

In the main text, we have presented the results for $ {\langle \hat O \rangle }_{A_4 A_4}$ which we report here as well for concreteness, 
\ba
\label{B-AA}
{\langle \hat O \rangle }_{A_4 A_4}= \int_{-\infty}^{\tau_0} d \tau_2 \int_{-\infty}^{\tau_2} d \tau_1 A_4(\tau_1) A_4(\tau_2)  \int d^3 \bfx \int d^3  {\bf y}
\Big \langle  \Big[ \calR^2 \calR'^2 (\bfx, \tau_1) ,  \big[ \hat O (\tau_0)  ,  \calR^2 \calR'^2  ({\bf{y}}, \tau_2) \big] \Big]
 \Big \rangle \, .
\ea
Going to the Fourier space and  after performing the contractions and imposing the constraints from the delta functions, in the soft limit where $p\ll q, k$, we obtain
\ba
{\big\langle \calR_{\bfp_1} \calR_{\bfp_2}(\tau_0) \big\rangle }_{A_4 A_4}= (2 \pi)^3 \delta^3(\bfp_1 + \bfp_2)
\int_{-\infty}^{\tau_0} d\tau_2
 \int_{-\infty}^{\tau_2} d\tau_1 A_4(\tau_1) A_4( \tau_2) 
 \int \frac{d^3 \bfq}{(2 \pi)^3} \int \frac{d^3 \bfk}{(2 \pi)^3} 
 F(\tau_1, \tau_2; k, q),  \nonumber
\ea
in which the function $F(\tau_1, \tau_2; k, q)$ has the following 9 contributions, 
\ba
\label{a-term}
&&
(a_0):  (-4) (2)^2 \im \big[  \calR_p(\tau_0)^2  {\calR_p^{*}(\tau_2) }^2   \big]
\im\big[ {\calR_k'(\tau_1)}^2 {\calR_k^{'*}(\tau_2)}^2  \big] |\calR_q(\tau_1)|^2
\nonumber\\
&&
\label{b-term}
(b_0):  (-4) (2)^2 \im \big[  \calR_p(\tau_0)^2  {\calR_p^{*}(\tau_2) }^2   \big]
\im\big[ {\calR_k(\tau_1)}^2 {\calR_k^{'*}(\tau_2)}^2  \big] |\calR'_q(\tau_1)|^2
\nonumber\\
&&
\label{c-term}
(c_0):  (-4) (2)^3 \im \big[  \calR_p(\tau_0)^2  {\calR_p^{*}(\tau_2) }^2   \big]
\im\big[ {\calR_k(\tau_1)} {\calR_k'(\tau_1)} {\calR_k^{'*}(\tau_2)}^2 \big]
\mathrm{Re} \big[\calR_q(\tau_1) \calR'_q(\tau_1)^*   \big],
\ea
and
\ba
\label{d-term}
&&
(d_0):  (-4) (2)^4 \im \big[  \calR_p(\tau_0)^2  {\calR_p^{*}(\tau_2) } {\calR_p^{'*}(\tau_2) }   \big]
\im\big[ {\calR_k'(\tau_1)}^2 {\calR_k^{'*}(\tau_2)} {\calR_k^{*}(\tau_2)}  \big] |\calR_q(\tau_1)|^2
\nonumber\\
&&
\label{e-term}
(e_0):  (-4) (2)^4  \im \big[  \calR_p(\tau_0)^2  {\calR_p^{*}(\tau_2) } {\calR_p^{'*}(\tau_2) }   \big]
\im\big[ {\calR_k(\tau_1)}^2 {\calR_k^{'*}(\tau_2)} {\calR_k^{*}(\tau_2)}  \big] |\calR'_q(\tau_1)|^2
\\
&&
\label{f-term}
(f_0):  (-4) (2)^5  \im \big[  \calR_p(\tau_0)^2  {\calR_p^{*}(\tau_2) } {\calR_p^{'*}(\tau_2) }   \big]
\im\big[ {\calR_k(\tau_1)} {\calR_k'(\tau_1)} {\calR_k^{'*}(\tau_2)}  {\calR_k^{*}(\tau_2)} \big]
\mathrm{Re} \big[\calR_q(\tau_1) \calR'_q(\tau_1)^*   \big],  \nonumber
\ea
and
\ba
\label{m-term}
&&
(m):  (-4) (2)^2 \im \big[  \calR_p(\tau_0)^2  {\calR_p^{'*}(\tau_2) }^2   \big]
\im\big[ {\calR_k'(\tau_1)}^2 {\calR_k^{*}(\tau_2)}^2  \big] |\calR_q(\tau_1)|^2,
\nonumber\\
&&
\label{n-term}
(n):  (-4) (2)^2 \im \big[  \calR_p(\tau_0)^2  {\calR_p^{'*}(\tau_2) }^2   \big]
\im\big[ {\calR_k(\tau_1)}^2 {\calR_k^{*}(\tau_2)}^2  \big] |\calR'_q(\tau_1)|^2,
\nonumber\\
&&
\label{r-term}
(r):  (-4) (2)^3 \im \big[  \calR_p(\tau_0)^2  {\calR_p^{'*}(\tau_2) }^2   \big]
\im\big[ {\calR_k(\tau_1)} {\calR_k'(\tau_1)} {\calR_k^{*}(\tau_2)}^2  \big]
\mathrm{Re} \big[\calR_q(\tau_1) \calR'_q(\tau_1)^*   \big] \, .
\ea

Now we consider the contribution ${\langle \hat O \rangle}_{A_4 B_4}$, yielding 
\ba
\label{B-AB}
{\langle \hat O \rangle }_{A_4 B_4}  =  \int_{-\infty}^{\tau_0} d \tau_2 \int_{-\infty}^{\tau_2} d \tau_1  A_4(\tau_1) B_4(\tau_2) 
\Big[ \prod_i^4 \int \frac{d^3 \bfq_i}{ (2 \pi)^3} (2 \pi)^3 \delta^3( \sum_i \bfq_i)\Big] 
\Big[\prod_j^4 \int \frac{d^3 \bfk_i}{(2 \pi)^3} (2 \pi)^3 \delta^3( \sum_i \bfk_i)
\Big]  \hspace{-1cm} \nonumber\\
 \times  \bigg \langle   \bigg[   \big( \hat\calR_{\bfq_1}   \hat\calR_{\bfq_2} \hat\calR'_{\bfq_3} \hat\calR'_{\bfq_4} \big) (\tau_1), \Big[
 \big( \hat\calR_{\bfp_1} \hat\calR_{\bfp_2}  \big) (\tau_0)  , \big( \hat\calR_{\bfk_1}  \hat\calR_{\bfk_2} \hat\calR_{\bfk_3}   \hat\calR_{\bfk_4} \big) (\tau_2)  \Big]  \bigg]
  \bigg \rangle i^2 \bfk_3\cdot \bfk_4 \, .
\ea
The leading contributions are those in which $\hat\calR_{\bfp}  \big (\tau_0)$
does not contract with $\hat\calR_{\bfk_3}(\tau_2)$ and $   \hat\calR_{\bfk_4}  (\tau_2)$. Correspondingly, the leading terms will be similar to terms 
$(a_0), (b_0)$ and $(c_0)$ in the analysis of ${\langle \hat O \rangle}_{A_4 A_4}$. More specifically, the leading terms are,
\ba
\label{a1-term}
&&
(a_1):  (-4) (2)^2 \im \big[  \calR_p(\tau_0)^2  {\calR_p^{*}(\tau_2) }^2   \big]
\im\big[ {\calR_k'(\tau_1)}^2 {\calR_k^{*}(\tau_2)}^2  \big] |\calR_q(\tau_1)|^2
(-i^2 k^2)
\nonumber\\
&&
\label{b1-term}
(b_1):  (-4) (2)^2 \im \big[  \calR_p(\tau_0)^2  {\calR_p^{*}(\tau_2) }^2   \big]
\im\big[ {\calR_k(\tau_1)}^2 {\calR_k^{*}(\tau_2)}^2  \big] |\calR'_q(\tau_1)|^2
(-i^2 k^2)
\nonumber\\
&&
\label{c1-term}
(c_1):  (-4) (2)^3 \im \big[  \calR_p(\tau_0)^2  {\calR_p^{*}(\tau_2) }^2   \big]
\im\big[ {\calR_k(\tau_1)} {\calR_k'(\tau_1)} {\calR_k^{*}(\tau_2)}^2 \big ]
\mathrm{Re} \big[\calR_q(\tau_1) \calR'_q(\tau_1)^*   \big] (-i^2 k^2).
\ea

Now consider ${\langle \hat O \rangle}_{B_4 A_4}$, yielding, 
\ba
\label{B-BA}
{\langle \hat O \rangle }_{B_4 A_4}  =  \int_{-\infty}^{\tau_0} d \tau_2 \int_{-\infty}^{\tau_2} d \tau_1  B_4(\tau_1) A_4(\tau_2) 
\Big[ \prod_i^4 \int \frac{d^3 \bfq_i}{ (2 \pi)^3} (2 \pi)^3 \delta^3( \sum_i \bfq_i)\Big] 
\Big[\prod_j^4 \int \frac{d^3 \bfk_i}{(2 \pi)^3} (2 \pi)^3 \delta^3( \sum_i \bfk_i)
\Big]  \hspace{-1cm} \nonumber\\
 \times  \bigg \langle   \bigg[   \big( \hat\calR_{\bfq_1}   \hat\calR_{\bfq_2} \hat\calR_{\bfq_3} \hat\calR_{\bfq_4} \big) (\tau_1), \Big[
 \big( \hat\calR_{\bfp_1} \hat\calR_{\bfp_2}  \big) (\tau_0)  , \big( \hat\calR_{\bfk_1}  \hat\calR_{\bfk_2} \hat\calR'_{\bfk_3}   \hat\calR'_{\bfk_4} \big) (\tau_2)  \Big]  \bigg]
  \bigg \rangle i^2 \bfq_3\cdot \bfq_4 \, .
\ea
The leading contributions are like $(a_0), (b_0), (c_0), (d_0), (e_0)$ and $(f_0)$ in  ${\langle \hat O \rangle}_{A_4 A_4}$, yielding
\ba
\label{a2-term}
&&
(a_2):  (-4) (2)^2 \im \big[  \calR_p(\tau_0)^2  {\calR_p^{*}(\tau_2) }^2   \big]
\im\big[ {\calR_k(\tau_1)}^2 {\calR_k^{'*}(\tau_2)}^2  \big] |\calR_q(\tau_1)|^2
(-i^2 k^2)
\nonumber\\
&&
\label{b2-term}
(b_2):  (-4) (2)^2 \im \big[  \calR_p(\tau_0)^2  {\calR_p^{*}(\tau_2) }^2   \big]
\im\big[ {\calR_k(\tau_1)}^2 {\calR_k^{'*}(\tau_2)}^2  \big] |\calR_q(\tau_1)|^2
(-i^2 q^2)
\nonumber\\
&&
\label{c2-term}
(c_2):  (-4) (2)^4 \im \big[  \calR_p(\tau_0)^2  {\calR_p^{*}(\tau_2) }^2   \big]
\im\big[ {\calR_k(\tau_1)}^2  {\calR_k^{'*}(\tau_2)}^2 \big]  |\calR_q(\tau_1)|^2
  (i^2 \bfk\cdot \bfq)
\ea
and
\ba
\label{d2-term}
&&
(d_2):  (-4) (2)^4 \im \big[  \calR_p(\tau_0)^2  {\calR_p^{*}(\tau_2) } {\calR_p^{'*}(\tau_2) }   \big]
\im\big[ {\calR_k(\tau_1)}^2 {\calR_k^{'*}(\tau_2)} {\calR_k^{*}(\tau_2)}  \big] |\calR_q(\tau_1)|^2 (-i^2 k^2)
\nonumber\\
&&
\label{e2-term}
(e_2):  (-4) (2)^4  \im \big[  \calR_p(\tau_0)^2  {\calR_p^{*}(\tau_2) } {\calR_p^{'*}(\tau_2) }   \big]
\im\big[ {\calR_k(\tau_1)}^2 {\calR_k^{'*}(\tau_2)} {\calR_k^{*}(\tau_2)}  \big] |\calR_q(\tau_1)|^2 (-i^2 q^2)
\nonumber\\
&&
\label{f2-term}
(f_2):  (-4) (2)^6  \im \big[  \calR_p(\tau_0)^2  {\calR_p^{*}(\tau_2) } {\calR_p^{'*}(\tau_2) }   \big]
\im\big[ {\calR_k(\tau_1)}^2 {\calR_k^{'*}(\tau_2)}  {\calR_k^{*}(\tau_2)}  \big]  
\calR_q(\tau_1)|^2 (i^2 \bfk \cdot \bfq)
\ea
From the 6 terms above, one can check that the terms $(c_2)$ and $(f_2)$ make zero contributions after performing the double momentum integrals of the form 
$\int d^3{\bfq} \, d^3{\bfk} \,  (\bfq \cdot \bfk) {\cal F}(\tau_1, \tau_2; k, q)$ which vanishes. 

Finally, considering ${\langle \hat O \rangle}_{B_4 B_4}$, we have 
\ba
\label{B-BB}
{\langle \hat O \rangle }_{B_4 B_4}  =  \int_{-\infty}^{\tau_0} d \tau_2 \int_{-\infty}^{\tau_2} d \tau_1  B_4(\tau_1) B_4(\tau_2) 
\Big[ \prod_i^4 \int \frac{d^3 \bfq_i}{ (2 \pi)^3} (2 \pi)^3 \delta^3( \sum_i \bfq_i)\Big] 
\Big[\prod_j^4 \int \frac{d^3 \bfk_i}{(2 \pi)^3} (2 \pi)^3 \delta^3( \sum_i \bfk_i)
\Big]  \hspace{-1cm} \nonumber\\
 \times  \bigg \langle   \bigg[   \big( \hat\calR_{\bfq_1}   \hat\calR_{\bfq_2} \hat\calR_{\bfq_3} \hat\calR_{\bfq_4} \big) (\tau_1), \Big[
 \big( \hat\calR_{\bfp_1} \hat\calR_{\bfp_2}  \big) (\tau_0)  , \big( \hat\calR_{\bfk_1}  \hat\calR_{\bfk_2} \hat\calR_{\bfk_3}   \hat\calR_{\bfk_4} \big) (\tau_2)  \Big]  \bigg]
  \bigg \rangle  ( \bfq_3\cdot \bfq_4) (  \bfk_3\cdot \bfk_4) \, .
\ea

The leading contributions are like $(a_0)$ and $ (b_0)$ terms, yielding 
\ba
\label{a3-term}
&&
(a_3):  (-4) (2)^2 \im \big[  \calR_p(\tau_0)^2  {\calR_p^{*}(\tau_2) }^2   \big]
\im\big[ {\calR_k(\tau_1)}^2 {\calR_k^{*}(\tau_2)}^2  \big] |\calR_q(\tau_1)|^2
(-i^2 k^2)^2
\nonumber\\
&&
\label{b3-term}
(b_3):  (-4) (2)^2 \im \big[  \calR_p(\tau_0)^2  {\calR_p^{*}(\tau_2) }^2   \big]
\im\big[ {\calR_k(\tau_1)}^2 {\calR_k^{*}(\tau_2)}^2  \big] |\calR_q(\tau_1)|^2
(-i^2 k^2)(-i^2 q^2)
\ea

{So in total, we have 15 leading terms (i.e. containing $p^{-3}$) given by 
$(a_0), (b_0), (c_0), (d_0), (e_0),$\linebreak  $(f_0)$, $(a_1), (b_1),  (c_1),  (a_2), (b_2), (d_2), (e_2), (a_3)$ and $(b_3)$ listed above. 
Out of these 15 contributions, only the terms $(b_0), (c_0), (d_0), (e_0), (f_0)$ scale likes $\Delta N^2 e^{12 \Delta N}$ while the remaining terms scale like 
$\Delta N e^{12 \Delta N}$. In the limit of large enough $\Delta N$, we may neglect the latter contributions as the subleading corrections. }
 
Calculating the leading terms from $(b_0), (c_0), (d_0), (e_0), (f_0)$, and neglecting the common factor $(2 \pi)^3 \delta^3(\bfp_1 + \bfp_2)$ and 
$\frac{(4 \pi)^2}{(2 \pi)^6}$ from the double azimuthal integrals over momentum,  we obtain,
\ba
\label{b0term}
&(b_0)&: \frac{27 H^6}{512 M_P^6 \epsilon_i^3 h p^3} (h^2 + 28 h - 384) \Delta N^2 e^{12 \Delta N} + {\cal O} (\Delta N) \, ,\\
\label{c0term}
&(c_0)&: \frac{-27 H^6}{64 M_P^6 \epsilon_i^3 h p^3} (h^2 + 8 h +96) \Delta N^2 e^{12 \Delta N} + {\cal O} (\Delta N)\, , \\
\label{d0term}
&(d_0)&: \frac{9 H^6}{256 M_P^6 \epsilon_i^3 h p^3} (h^2 + 8 h +48) \Delta N^2 e^{12 \Delta N} + {\cal O} (\Delta N)\, , \\
\label{e0term}
&(e_0)&: \frac{-9 H^6}{256 M_P^6 \epsilon_i^3 h p^3} (h^2 + 16 h +96) \Delta N^2 e^{12 \Delta N} + {\cal O} (\Delta N)\, , \\
\label{e0term}
&(f_0)&: \frac{-9 H^6}{16 M_P^6 \epsilon_i^3 h p^3} (h+6) \Delta N^2 e^{12 \Delta N} + {\cal O} (\Delta N)\, .
\ea

In comparison, we also present the remaining ten subleading terms $(a_0), (a_1), ..., (b_3)$ as well,
\ba
\label{a0term}
&(a_0)&: \frac{3 H^6 \Delta N e^{12 \Delta N}}{5600 M_P^6 \epsilon_i^3 h p^3} (346h-2997)  + {\cal O} (\Delta N^0) \, ,\\
\label{a1term}
&(a_1)&: \frac{ H^6 \Delta N e^{12 \Delta N}}{206976000 M_P^6 \epsilon_i^3 h p^3} (808500 h^2+51180091 h-44241120)  + {\cal O} (\Delta N^0)\, ,\\
\label{b1term}
&(b_1)&: \frac{3 H^6 \Delta N e^{12 \Delta N}}{71680 M_P^6 \epsilon_i^3 h p^3} (1544h-33024)  + {\cal O} (\Delta N^0) \, ,\\
\label{c1term}
&(c_1)&: \frac{ 3H^6 \Delta N e^{12 \Delta N}}{179200 M_P^6 \epsilon_i^3 h p^3} ( 2100 h^2+82832 h-344832)  + {\cal O} (\Delta N^0)\, ,\\
\label{a2term}
&(a_2)&: \frac{ H^6 \Delta N e^{12 \Delta N}}{5017600 M_P^6 \epsilon_i^3 h p^3} (78400 h^2+1628972 h-21780864)  + {\cal O} (\Delta N^0) \, ,\\
\label{b2term}
&(b2)&: \frac{ 9 H^6 \Delta N e^{12 \Delta N}}{17920 M_P^6 \epsilon_i^3 h p^3} (35 h^2-246 h-10368)  + {\cal O} (\Delta N^0)\, ,\\
\label{d2term}
&(d2)&: \frac{3 H^6 \Delta N e^{12 \Delta N}}{5017600 M_P^6 \epsilon_i^3  p^3} (4900 h^2-162656h-1374912)  + {\cal O} (\Delta N^0) \, ,\\
\label{e2term}
&(e2)&: \frac{ -3H^6 \Delta N e^{12 \Delta N}}{20070400 M_P^6 \epsilon_i^3  p^3} ( 4900 h^2-178080 h+5806080)  + {\cal O} (\Delta N^0)\, ,\\
\label{a3term}
&(a3)&: \frac{ H^6 \Delta N e^{12 \Delta N}}{3763200 M_P^6 \epsilon_i^3  p^3} (478062 h-3175200)  + {\cal O} (\Delta N^0) \, ,\\
\label{b3term}
&(b3)&: \frac{ H^6 \Delta N e^{12 \Delta N}}{313600 M_P^6 \epsilon_i^3  p^3} ( 58637 h-423360)  + {\cal O} (\Delta N^0)\,.
\ea

Combining the five leading contributions $(b_0), (c_0), (d_0), (e_0), (f_0)$, and including all numerical factors,  
we obtain our final result,
\ba
\label{final}
\label{e0term}
{\big\langle \calR_{\bfp_1} \calR_{\bfp_2} \big\rangle } \big |_\mathrm{2-loops}
= (2 \pi)^3 \delta^3(\bfp_1 + \bfp_2)
  \frac{-27 H^6 (4 \pi)^2}{512 M_P^6 \epsilon_i^3 h p^3 (2 \pi)^6} (23 h^2 + 132 h + 1152) \Delta N^2 e^{12 \Delta N} + {\cal O} (\Delta N)\, . \nonumber
\ea
Now, multiplying by the factor $\frac{p^3}{2 \pi^2}$ to construct the dimensionless power spectrum, we end up with our fractional two-loop correction as follows,
\ba
\label{two-loop2}
\frac{\Delta \calP^{(2-\mathrm{loop})}}{\calP_{{\mathrm{CMB}}}}\simeq  -\frac{27(23 h^2 + 132 h + 1152)}{8 h}\,  e^{12 \Delta N} N^2 \calP_{{\mathrm{CMB}}}^2  + {\cal O} (\Delta N)\, ,
\ea
where $\calP_{{\mathrm{CMB}}}$ is the tree-level  power spectrum for the CMB scale modes given in Eq. (\ref{P-CMB}).


{}


\begin{thebibliography}{}


\bibitem{Kristiano:2022maq}
J.~Kristiano and J.~Yokoyama,
Phys. Rev. Lett. \textbf{132}, no.22, 221003 (2024), 
[arXiv:2211.03395 [hep-th]].

\bibitem{Kristiano:2023scm}
J.~Kristiano and J.~Yokoyama,
Phys. Rev. D \textbf{109}, no.10, 103541 (2024),
[arXiv:2303.00341 [hep-th]].


\bibitem{Riotto:2023hoz}
A.~Riotto,
[arXiv:2301.00599 [astro-ph.CO]].

\bibitem{Riotto:2023gpm}
A.~Riotto,
[arXiv:2303.01727 [astro-ph.CO]].

\bibitem{Choudhury:2023vuj}
S.~Choudhury, M.~R.~Gangopadhyay and M.~Sami,
Eur. Phys. J. C \textbf{84}, no.9, 884 (2024),
[arXiv:2301.10000 [astro-ph.CO]].

\bibitem{Choudhury:2023jlt}
S.~Choudhury, S.~Panda and M.~Sami,
Phys. Lett. B \textbf{845}, 138123 (2023),
[arXiv:2302.05655 [astro-ph.CO]].

\bibitem{Choudhury:2023rks}
S.~Choudhury, S.~Panda and M.~Sami,
JCAP \textbf{11}, 066 (2023),
[arXiv:2303.06066 [astro-ph.CO]].

\bibitem{Choudhury:2023hvf}
S.~Choudhury, S.~Panda and M.~Sami,
JCAP \textbf{08}, 078 (2023), 
[arXiv:2304.04065 [astro-ph.CO]].

\bibitem{Choudhury:2024one}
S.~Choudhury, A.~Karde, S.~Panda and M.~Sami,
JCAP \textbf{07}, 034 (2024),
[arXiv:2401.10925 [astro-ph.CO]].

\bibitem{Choudhury:2024aji}
S.~Choudhury and M.~Sami,
[arXiv:2407.17006 [gr-qc]].

\bibitem{Firouzjahi:2023aum}
H.~Firouzjahi,
JCAP \textbf{10}, 006 (2023), 
[arXiv:2303.12025 [astro-ph.CO]].
  
\bibitem{Motohashi:2023syh}
H.~Motohashi and Y.~Tada,
JCAP \textbf{08}, 069 (2023),
[arXiv:2303.16035 [astro-ph.CO]].

\bibitem{Firouzjahi:2023ahg}
H.~Firouzjahi and A.~Riotto,
JCAP \textbf{02}, 021 (2024),
[arXiv:2304.07801 [astro-ph.CO]].

\bibitem{Tasinato:2023ukp}
G.~Tasinato,
Phys. Rev. D \textbf{108}, no.4, 043526 (2023), 
[arXiv:2305.11568 [hep-th]].



\bibitem{Franciolini:2023agm}
G.~Franciolini, A.~Iovino, Junior., M.~Taoso and A.~Urbano,
Phys. Rev. D \textbf{109}, no.12, 123550 (2024), 
[arXiv:2305.03491 [astro-ph.CO]].

\bibitem{Firouzjahi:2023btw}
H.~Firouzjahi,
Phys. Rev. D \textbf{108}, no.4, 043532 (2023), 
[arXiv:2305.01527 [astro-ph.CO]].

\bibitem{Maity:2023qzw}
S.~Maity, H.~V.~Ragavendra, S.~K.~Sethi and L.~Sriramkumar,
JCAP \textbf{05}, 046 (2024),
[arXiv:2307.13636 [astro-ph.CO]].

\bibitem{Cheng:2023ikq}
S.~L.~Cheng, D.~S.~Lee and K.~W.~Ng,
JCAP \textbf{03}, 008 (2024),
[arXiv:2305.16810 [astro-ph.CO]].

\bibitem{Fumagalli:2023loc}
J.~Fumagalli, S.~Bhattacharya, M.~Peloso, S.~Renaux-Petel and L.~T.~Witkowski,
JCAP \textbf{04}, 029 (2024), 
[arXiv:2307.08358 [astro-ph.CO]].

\bibitem{Nassiri-Rad:2023asg}
A.~Nassiri-Rad and K.~Asadi,
JCAP \textbf{04}, 009 (2024), 
[arXiv:2310.11427 [astro-ph.CO]].

\bibitem{Meng:2022ixx}
D.~S.~Meng, C.~Yuan and Q.~g.~Huang,
Phys. Rev. D \textbf{106}, no.6, 063508 (2022). 




\bibitem{Cheng:2021lif}
S.~L.~Cheng, D.~S.~Lee and K.~W.~Ng,
Phys. Lett. B \textbf{827}, 136956 (2022). 


\bibitem{Fumagalli:2023hpa}
J.~Fumagalli,
[arXiv:2305.19263 [astro-ph.CO]].

\bibitem{Tada:2023rgp}
Y.~Tada, T.~Terada and J.~Tokuda,
JHEP \textbf{01}, 105 (2024), 
[arXiv:2308.04732 [hep-th]].


\bibitem{Firouzjahi:2023bkt}
H.~Firouzjahi,
Phys. Rev. D \textbf{109}, no.4, 043514 (2024), 
[arXiv:2311.04080 [astro-ph.CO]].

\bibitem{Iacconi:2023slv}
L.~Iacconi and D.~J.~Mulryne,
JCAP \textbf{09}, 033 (2023). 


\bibitem{Davies:2023hhn}
M.~W.~Davies, L.~Iacconi and D.~J.~Mulryne,
JCAP \textbf{04}, 050 (2024),
[arXiv:2312.05694 [astro-ph.CO]].


\bibitem{Iacconi:2023ggt}
L.~Iacconi, D.~Mulryne and D.~Seery,
JCAP \textbf{06}, 062 (2024),
[arXiv:2312.12424 [astro-ph.CO]].


\bibitem{Kristiano:2024vst}
J.~Kristiano and J.~Yokoyama,
JCAP \textbf{10}, 036 (2024), 
[arXiv:2405.12145 [astro-ph.CO]].

\bibitem{Kristiano:2024ngc}
J.~Kristiano and J.~Yokoyama,
[arXiv:2405.12149 [astro-ph.CO]].

\bibitem{Ballesteros:2024zdp}
G.~Ballesteros and J.~G.~Egea,
JCAP \textbf{07}, 052 (2024), 
[arXiv:2404.07196 [astro-ph.CO]].

\bibitem{Kawaguchi:2024lsw}
R.~Kawaguchi, S.~Tsujikawa and Y.~Yamada,
Phys. Lett. B \textbf{856}, 138962 (2024),
[arXiv:2403.16022 [hep-th]].

\bibitem{Braglia:2024zsl}
M.~Braglia and L.~Pinol,
JHEP \textbf{08}, 068 (2024),
[arXiv:2403.14558 [astro-ph.CO]].


\bibitem{Firouzjahi:2024psd}
H.~Firouzjahi,
Phys. Rev. D \textbf{110}, no.4, 043519 (2024), 
[arXiv:2403.03841 [astro-ph.CO]].

\bibitem{Caravano:2024moy}
A.~Caravano, G.~Franciolini and S.~Renaux-Petel,
[arXiv:2410.23942 [astro-ph.CO]].

\bibitem{Caravano:2024tlp}
A.~Caravano, K.~Inomata and S.~Renaux-Petel,
Phys. Rev. Lett. \textbf{133}, no.15, 15 (2024), 
[arXiv:2403.12811 [astro-ph.CO]].

\bibitem{Saburov:2024und}
S.~Saburov and S.~V.~Ketov,
Universe \textbf{10}, no.9, 354 (2024), 
[arXiv:2402.02934 [gr-qc]].


\bibitem{Seery:2007wf}
D.~Seery,
JCAP \textbf{02}, 006 (2008), 
[arXiv:0707.3378 [astro-ph]].

\bibitem{Seery:2007we}
D.~Seery,
JCAP \textbf{11}, 025 (2007), 
[arXiv:0707.3377 [astro-ph]].


\bibitem{Senatore:2009cf}
L.~Senatore and M.~Zaldarriaga,
JHEP \textbf{12}, 008 (2010), 
[arXiv:0912.2734 [hep-th]].

\bibitem{Pimentel:2012tw}
G.~L.~Pimentel, L.~Senatore and M.~Zaldarriaga,
JHEP \textbf{07}, 166 (2012). 

\bibitem{Inomata:2022yte}
K.~Inomata, M.~Braglia, X.~Chen and S.~Renaux-Petel,
JCAP \textbf{04}, 011 (2023), 
[erratum: JCAP \textbf{09}, E01 (2023)], 
[arXiv:2211.02586 [astro-ph.CO]].




\bibitem{Ivanov:1994pa}
P.~Ivanov, P.~Naselsky and I.~Novikov,
Phys. Rev. D \textbf{50}, 7173-7178 (1994). 

\bibitem{Garcia-Bellido:2017mdw}
J.~Garcia-Bellido and E.~Ruiz Morales,
Phys. Dark Univ. \textbf{18}, 47-54 (2017). 

\bibitem{Germani:2017bcs}
C.~Germani and T.~Prokopec,
Phys. Dark Univ. \textbf{18}, 6-10 (2017). 




\bibitem{Biagetti:2018pjj}
M.~Biagetti, G.~Franciolini, A.~Kehagias and A.~Riotto,
JCAP \textbf{07}, 032 (2018). 

\bibitem{Khlopov:2008qy}
M.~Y.~Khlopov,
Res. Astron. Astrophys. \textbf{10}, 495-528 (2010), 
[arXiv:0801.0116 [astro-ph]].

\bibitem{Ozsoy:2023ryl}
O.~\"Ozsoy and G.~Tasinato,
Universe \textbf{9}, no.5, 203 (2023),
[arXiv:2301.03600 [astro-ph.CO]].


\bibitem{Byrnes:2021jka}
C.~T.~Byrnes and P.~S.~Cole,
[arXiv:2112.05716 [astro-ph.CO]].

\bibitem{Escriva:2022duf}
A.~Escriv\`a, F.~Kuhnel and Y.~Tada,
[arXiv:2211.05767 [astro-ph.CO]].


\bibitem{Pi:2024jwt}
S.~Pi,
[arXiv:2404.06151 [astro-ph.CO]].

\bibitem{Maldacena:2002vr} 
  J.~M.~Maldacena,
  JHEP {\bf 0305}, 013 (2003),
  [astro-ph/0210603].

\bibitem{Jarnhus:2007ia}
P.~R.~Jarnhus and M.~S.~Sloth,
JCAP \textbf{02}, 013 (2008), 
[arXiv:0709.2708 [hep-th]].

\bibitem{Arroja:2008ga}
F.~Arroja and K.~Koyama,
Phys. Rev. D \textbf{77}, 083517 (2008), 
[arXiv:0802.1167 [hep-th]].


\bibitem{Inomata:2024lud}
K.~Inomata,
Phys. Rev. Lett. \textbf{133}, no.14, 141001 (2024), 
[arXiv:2403.04682 [astro-ph.CO]].

\bibitem{Kawaguchi:2024rsv}
R.~Kawaguchi, S.~Tsujikawa and Y.~Yamada,
[arXiv:2407.19742 [hep-th]].


\bibitem{Fumagalli:2024jzz}
J.~Fumagalli,
[arXiv:2408.08296 [astro-ph.CO]].






\bibitem{Kinney:2005vj} 
  W.~H.~Kinney,
  Phys.\ Rev.\ D {\bf 72}, 023515 (2005),
  [gr-qc/0503017].
  

\bibitem{Namjoo:2012aa} 
  M.~H.~Namjoo, H.~Firouzjahi and M.~Sasaki,
  Europhys.\ Lett.\  {\bf 101}, 39001 (2013).

\bibitem{Martin:2012pe}
J.~Martin, H.~Motohashi and T.~Suyama,
Phys. Rev. D \textbf{87}, no.2, 023514 (2013). 
  
  
\bibitem{Chen:2013aj} 
  X.~Chen, H.~Firouzjahi, M.~H.~Namjoo and M.~Sasaki,
  Europhys.\ Lett.\  {\bf 102}, 59001 (2013).
   
\bibitem{Morse:2018kda}
M.~J.~P.~Morse and W.~H.~Kinney,
Phys. Rev. D \textbf{97}, no.12, 123519 (2018). 

\bibitem{Lin:2019fcz}
W.~C.~Lin, M.~J.~P.~Morse and W.~H.~Kinney,
JCAP \textbf{09}, 063 (2019).

\bibitem{Dimopoulos:2017ged}
K.~Dimopoulos,
Phys. Lett. B \textbf{775}, 262-265 (2017), 
[arXiv:1707.05644 [hep-ph]].
  
  
\bibitem{Chen:2013eea} 
  X.~Chen, H.~Firouzjahi, E.~Komatsu, M.~H.~Namjoo and M.~Sasaki,
  JCAP {\bf 1312}, 039 (2013).
  
\bibitem{Akhshik:2015nfa}
M.~Akhshik, H.~Firouzjahi and S.~Jazayeri,
JCAP \textbf{07}, 048 (2015), 
[arXiv:1501.01099 [hep-th]].
  
\bibitem{Akhshik:2015rwa}
M.~Akhshik, H.~Firouzjahi and S.~Jazayeri,
JCAP \textbf{12}, 027 (2015), 
[arXiv:1508.03293 [hep-th]].


\bibitem{Mooij:2015yka}
S.~Mooij and G.~A.~Palma,
JCAP \textbf{11}, 025 (2015), 
[arXiv:1502.03458 [astro-ph.CO]].

\bibitem{Bravo:2017wyw}
R.~Bravo, S.~Mooij, G.~A.~Palma and B.~Pradenas,
JCAP \textbf{05}, 024 (2018). 

\bibitem{Finelli:2017fml}
B.~Finelli, G.~Goon, E.~Pajer and L.~Santoni,
Phys. Rev. D \textbf{97}, no.6, 063531 (2018). 
 
\bibitem{Passaglia:2018ixg}
S.~Passaglia, W.~Hu and H.~Motohashi,
Phys. Rev. D \textbf{99}, no.4, 043536 (2019). 

\bibitem{Pi:2022ysn}
S.~Pi and M.~Sasaki,
Phys. Rev. Lett. \textbf{131}, no.1, 011002 (2023). 

\bibitem{Ozsoy:2021pws}
O.~\"Ozsoy and G.~Tasinato,
Phys. Rev. D \textbf{105}, no.2, 023524 (2022),


\bibitem{Firouzjahi:2023xke}
H.~Firouzjahi and A.~Riotto,
Phys. Rev. D \textbf{108}, no.12, 123504 (2023). 

\bibitem{Namjoo:2023rhq}
M.~H.~Namjoo,
JCAP \textbf{05}, 041 (2024), 
[arXiv:2311.12777 [astro-ph.CO]].


\bibitem{Namjoo:2024ufv}
M.~H.~Namjoo and B.~Nikbakht,
JCAP \textbf{08}, 005 (2024), 
[arXiv:2401.12958 [astro-ph.CO]].

\bibitem{Cai:2018dkf}
Y.~F.~Cai, X.~Chen, M.~H.~Namjoo, M.~Sasaki, D.~G.~Wang and Z.~Wang,
JCAP \textbf{05}, 012 (2018). 




\bibitem{Cheung:2007st} 
  C.~Cheung, P.~Creminelli, A.~L.~Fitzpatrick, J.~Kaplan and L.~Senatore,
  JHEP {\bf 0803}, 014 (2008).


\bibitem{Cheung:2007sv} 
  C.~Cheung, A.~L.~Fitzpatrick, J.~Kaplan and L.~Senatore,
  JCAP {\bf 0802}, 021 (2008).
  
  


\bibitem{Weinberg:2005vy}
S.~Weinberg,
Phys. Rev. D \textbf{72}, 043514 (2005)\, .
[arXiv:hep-th/0506236 [hep-th]].


\bibitem{Chen:2006dfn}
X.~Chen, M.~x.~Huang and G.~Shiu,
Phys. Rev. D \textbf{74}, 121301 (2006).

\bibitem{Chen:2009bc}
X.~Chen, B.~Hu, M.~x.~Huang, G.~Shiu and Y.~Wang,
JCAP \textbf{08}, 008 (2009).

\bibitem{Weinberg:1995mt}
S.~Weinberg,
Cambridge University Press, 2005.


\bibitem{Nikbakht}
Hassan Firouzjahi and Bahar Nikbakht,  work in progress. 



\end{thebibliography}
\end{document}